\documentclass[aps,reprint,prd,showpacs,amsmath,amssymb,nofootinbib]{revtex4-1}
\usepackage{graphicx}
\usepackage{color}
\usepackage{times}
\usepackage{inputenc}
\usepackage{bm}
\usepackage{multirow}
\usepackage{float}
\usepackage{url}
\usepackage{natbib}
\usepackage{tensor}
\usepackage[colorlinks=true,citecolor=blue,urlcolor=blue,linkcolor=red]{hyperref}
\usepackage[normalem]{ulem}

\def\Msun{\,M_{\odot}}
\def\fm3{\;\text{fm}^{-3}}
\def\mev{\;\text{MeV}}

\begin{document}
\title{Interacting $ud$ and $uds$ quark matter at finite densities and quark stars}

\author{Wen-Li Yuan$^{1,2}$}
\email{wlyuan@smail.nju.edu.cn}
\author{Ang Li$^{2}$}
\email[Corresponding author: ]{liang@xmu.edu.cn}
\author{Zhiqiang Miao$^{2}$}
\author{Bingjun Zuo$^{1}$}
\author{Zhan Bai$^{3}$}

\author{Wen-Li Yuan,$^{1,2}$ Ang Li,$^{2}$ Zhiqiang Miao,$^{2}$ Bingjun Zuo,$^{1}$ Zhan Bai$^{3}$}
\affiliation{$^1$Department of Physics, Nanjing University, Nanjing 210093, China;\\ 
$^2$Department of Astronomy, Xiamen University, Xiamen, Fujian 361005, China;\\ 
$^3$Institute of Theoretical Physics, Chinese Academy of Sciences, Beijing 100190, China 
}


\begin{abstract}
The stability and equation of state of quark matter are studied within both two-flavor and (2+1)-flavor Nambu-Jona-Lasinio (NJL) models including the vector interactions. 
 With a free parameter $\alpha$, the Lagrangian is constructed by two parts, the original NJL Lagrangian and the Fierz transformation of it, as $\mathcal{L}=(1-\alpha) \mathcal{L}_{\mathrm{NJL}}+\alpha \mathcal{L}_{\text {Fierz}}$.
We find that there is a possibility for both $ud$ nonstrange and $uds$ strange matter being absolute stable, depending on the interplay of the confinement with quark vector interaction and the exchange interaction channels.  
The calculated quark star properties can reconcile with the recently measured masses and radii of PSR J0030+0451 and PSR J0740+6620, as well as the tidal deformability of GW170817.
Furthermore, the more strongly-interacting quark matter in the nonstrange stars allows a stiffer equation of state and consequently a higher maximum mass ($\sim2.7\Msun$) than the strange ones ($\sim2.1\Msun$).
The sound velocities in strange and nonstrange quark star matter are briefly discussed compared to those of neutron star matter.
\end{abstract}

\maketitle 

\section{Introduction}

Thanks to the new generation of space and terrestrial facilities, the ever-increasing data from nuclear physics experiments and astrophysical observations have incited an intense research activity towards understanding the dense matter equation of state (EoS) and the composition of compact stars.
The long-standing important fundamental questions may have the possibility to be answered within a few decades.
One of the crucial questions related to the dense matter EoS is whether strange quark matter exists or not. After decades of speculation~\cite{1971PhRvD...4.1601B,1984PhRvD..30..272W}, it is still completely speculative, known as the Bodmer-Witten hypothesis.

Strange quark matter is a bulk quark matter phase consisting of $u$, $d$, and $s$ quarks in $\beta$-equilibrium in approximately equal proportions (plus a small fraction of electrons), with a density comparable to that of atomic nuclei.
Quark matter may exist in lumps ranging in size from a few fermis up to possible self-bound quark stars.
The possibility that ordinary atomic nuclei could be only a metastable state with respect to the true ground state of baryonic matter at zero pressure was first discussed by Bodmer (1971)~\cite{1971PhRvD...4.1601B}, and it was pointed out that the possible existence of 
$uds$ matter was not in conflict with the experimental data.
Quantitative studies~\cite{1979PhRvL..43.1292C,Terazawa1979,1984PhRvD..30..272W,1984PhRvD..30.2379F} were then carried out a few years after the Bodmer paper using the MIT bag model~\cite{1974PhRvD...9.3471C}.

In the bag model, all important quark interactions are assumed to be represented by the perturbative quantum chromodynamics (QCD) vacuum energy density $B$, namely the excess of the energy density of the QCD vacuum (inside the bag) over the energy density of the ordinary vacuum (outside the bag).
The energy density in the bag model is, therefore the sum of the bag constant, the kinetic energy of quarks and their interaction energy, and the interaction energy is usually 
calculated from the perturbative schemes of the QCD.
The EoS, namely the energy-density-pressure relation, is almost linear in the bag model~\cite{2000A&A...359..311Z}.
Since the perturbative QCD is inadequate for the treating of the quark matter EoS,
there have been numerous attempts to include non-perturbative effects in more advanced models, such as the Dyson-Schwinger equation approach~\cite{2000NuPhS..45S...1R,2016EPJA...52..291C,2021EPJC...81..612B}, the Nambu-Jona-Lasinio (NJL) model~(for review see Refs.~\cite{1991PrPNP..27..195V,1992RvMP...64..649K,1994PhR...247..221H,2005PhR...407..205B}), the density-dependent quark masses~\cite{1989PhLB..229..112C,1998PhLB..438..123D,1999PhRvC..61a5201P,2000PhRvC..62a5204W,2010MNRAS.402.2715L}, and the quark-meson coupling model~\cite{1988PhLB..200..235G,1996NuPhA.601..349G,1996PhLB..374...13J,1997NuPhA.626..966M,2016PhRvL.116i2501S,2020JHEAp..28...19L} as an incomplete list. According to the Bodmer-Witten hypothesis, i.e., 
the absolute stability of quark matter with strangeness,
the energy per baryon of $uds$ quark matter could be smaller than that of an $^{56}\rm Fe$ nucleus $E/A (P=0)=(56m_N-56\times8.8\mev)/56=930\mev$. 
Therefore all compact objects would be strange quark stars instead of neutron stars, despite the timescale for the conversion might be extremely long. The physics of strange quark matter is reviewed in e.g., Refs.~\cite{1996csnp.book.....G,1999LNP...516..162M,2005PrPNP..54..193W,2007ASSL..326.....H}.
Furthermore, although it was usually regarded that the $uds$ matter should always be energetically preferable over the $ud$ matter due to the extra Fermi well by the strange quarks, a recent study~\cite{2018PhRvL.120v2001H} showed that $ud$ matter could be in general more stable than $uds$ matter when taking the flavor-dependent feedback of the quark gas on the QCD vacuum into account.
In this case, $ud$ matter, instead of the originally-proposed $uds$ matter, would be the true ground state of cold, dense baryonic matter at zero pressure, i.e., absolutely stable (see e.g. Refs.~\cite{2019PhRvD.100d3018Z,2019PhRvD.100l3003W,2020arXiv200900942C,2020MPLA...3550321W,2021NuPhB.97115540X} for recent discussions of nonstrange quark stars). 

The NJL model is an important and valid effective quark theory, which serves as a suitable approximation to QCD in the low-energy and long-wavelength limit by assuming that gluon degrees of freedom can be frozen into effective point-like interactions between quarks. 
Its Lagrangian is constructed in such a way that the basic symmetries of QCD, which are observed in nature, are part and parcel of it. 
Moreover, the NJL model is found to work rather well in describing phenomenologically the interaction responsible for the quark flavor dynamics at intermediate energies~\cite{1992RvMP...64..649K,1994PhR...247..221H,2005PhR...407..205B}. 
Recently, to improve the description of the strong interaction matter at large chemical potential, and as an attempt to resolve the huge contradictions between the results drawn from the quark-gluon degrees of freedom and the expected results derived from the hadron degrees of freedom, Ref.~\cite{2019ChPhC..43h4102W}
proposed a modified NJL model containing both the original model Lagrangian and the Fierz transformation of it with the parameter $(1-\alpha)$ and $\alpha$ adjusting the weight of these two parts, respectively. 
Namely, $\alpha$ can be adjusted in the range of $0$ to $1$ to be consistent with finite-density constraints. It is a more general version of the NJL-like models compared to the specific one of $\alpha=0.5$ introduced in Ref.~\cite{1992RvMP...64..649K}.
The modified version of NJL model has been applied to investigating the color superconductivity~\cite{2020PhRvD.102e4028S}, the location of the QCD critical endpoint~\cite{2019PhRvD.100i4012Y}, as well as the QCD phase diagram at finite chemical potentials and finite temperature~\cite{2021ChPhC..45f4102W}. 
In the present work, we use the modified NJL model to study both $ud$ and $uds$ quark matter and the corresponding self-bound stars. 
For this we need further extend the model~\cite{2020PhRvD.101f3023L} to include the vector interactions shown as necessary for the study of dense stellar matter and compact stars~\cite{1998PhLB..438..123D,2012PhRvD..85k4017S,2013PhRvD..88h5001K,2015ApJ...810..134K,2017PhRvD..96h3019C,2018Univ....4...30C,2019JPhG...46c4002D,2020EPJST.229.3629O,2021Symm...13..124A,2021PhRvD.104h3011A}. 
This is also one of the first studies to evaluate systematically the stability of both strange and nonstrange quark matter, in connection with available multi-messenger stellar observations.

This paper is organized as follows.
In Section \ref{sec:form}, we introduce the two-flavor and (2+1)-flavor modified NJL models for describing the quark matter, including the vector interactions.
Section \ref{sec:star} discusses the results on quark matter EoS and quark stars, along with the observational constraints.
Our results are summarized in Section \ref{sec:summary}.

\section{Formalism}\label{sec:form}

In this section, we write down the NJL models to describe the effective interactions between quarks. As mentioned in the introduction, our calculations on interacting quark matter are done for both $ud$ quark matter and $uds$ quark matter.

\subsection{Two-flavor NJL model}
The Lagrangian of the two-flavor NJL model reads:
\begin{equation}
\mathcal{L}_{\mathrm{NJL}}^{~2f} =\mathcal{L}_{0}+\mathcal{L}_{\mathrm{int}}^{~2f} \ , \label{eq1}
\end{equation}
where $\mathcal{L}_{0} =\bar{\psi}\left(i \gamma^{\mu}\partial_{\mu} -m +\mu \gamma^{0} \right)\psi$ is the relativistic free (Dirac) Lagrangian which describes the propagation of non-interacting fermions. $\psi$ is the quark field operator with color, flavor, and Dirac indices.
$\mu$ is the flavor-dependent quark chemical potential. $m$ is the diagonal mass matrix for quarks in flavor space $m=\mathrm{diag}(m_u,m_d)$, which contains the small current quark masses and introduces a small explicit chiral symmetry breaking. Here, we take $m_u=m_d$.

The effective NJL-type interactions are four-fermion interactions, which simplify the gauge interaction coupling the quarks to the gluon dynamics in QCD.
The second term in Eq.~(\ref{eq1}) describes the four-fermion contact interactions between quarks, composing of scalar and vector interaction $\mathcal{L}_{\mathrm{int}}^{~2f} =\mathcal{L}_{\sigma}^{4} + \mathcal{L}_{V}^{4}$:
\begin{equation}
\mathcal{L}_{\sigma}^{4} =G\left[(\bar{\psi} \psi)^{2}+\left(\bar{\psi} i \gamma^{5} \tau \psi\right)^{2}\right] \ , \label{eq2}
\end{equation}
    \vspace{-0.4cm}
\begin{equation}
\mathcal{L}_{V}^{4} = -G_V(\bar{\psi} \gamma^{\mu}\psi)^{2} \ , \label{eq3}
\end{equation}
which emerge as the simplest way to write an interaction with only quark degrees of freedom that satisfies the flavor symmetries characterized by the group $S U(2)_{V} \times S U(2)_{A} \times U(1)_{B}$. The standard two-flavor NJL Lagrangian with interaction terms in the scalar and pseudoscalar channels is given by Eq.~(\ref{eq2}).
The Lagrangian $\mathcal{L}_{V}^{4}$ in Eq.~(\ref{eq3}) is the phenomenological vector interaction, which produces universal repulsion between quarks, and the finite-density environment brings a significant contribution to this channel.
The scalar and vector contact interaction coupling constants, $G$ and $G_{V}$ we consider here, can be interpreted to encode all the gluonic contribution of the strong interaction.

In the following, we further consider the effect of a rearrangement of fermion field operators. As a purely technical device to examine the exchange channels influence that occur in quartic products at the same space-time point~\cite{1992RvMP...64..649K,2005PhR...407..205B}, the Fierz identity of the four-fermion interactions in the two-flavor NJL model is
\begin{equation}
\mathcal{F}(\mathcal{L}_{\mathrm{int}}^{2f} )=\mathcal{F}(\mathcal{L}_{\sigma}^{4})+ \mathcal{F}(\mathcal{L}_{V}^{4}) \ ,\label{eq4}
\end{equation}
where
\begin{equation}
\begin{aligned}
\mathcal{F}(\mathcal{L}_{\sigma}^{4})=& \frac{G}{8 N_{c}}\left[2(\bar{\psi} \psi)^{2}+2\left(\bar{\psi} i \gamma^{5} \tau \psi\right)^{2}-2(\bar{\psi} \tau \psi)^{2}\right.\\
&-2\left(\bar{\psi} i \gamma^{5} \psi\right)^{2}-4\left(\bar{\psi} \gamma^{\mu} \psi\right)^{2}-4\left(\bar{\psi} i \gamma^{\mu} \gamma^{5} \psi\right)^{2} \\
&\left.+\left(\bar{\psi} \sigma^{\mu \nu} \psi\right)^{2}-\left(\bar{\psi} \sigma^{\mu \nu} \tau \psi\right)^{2}\right] \ , \label{eq5}
\end{aligned}
\end{equation}
and
\begin{equation}
\begin{aligned}
\mathcal{F}(\mathcal{L}_{V}^{4})= & \frac{G_V}{2N_{c}}\left[(\bar{\psi} \psi)^{2}+ (\bar{\psi}i\gamma^{5}\psi)^{2}-\frac{1}{2}(\bar{\psi} \gamma^{\mu} \psi)^{2}\right.\\
&-\frac{1}{2}(\bar{\psi} \gamma^{\mu} \gamma^{5} \psi)^{2}+ (\bar{\psi}\tau \psi)^{2} +(\bar{\psi}i\gamma^{5}\tau \psi)^{2}\\
&\left.-\frac{1}{2}(\bar{\psi} \gamma^{\mu}\tau \psi)^{2} -\frac{1}{2}(\bar{\psi} \gamma^{\mu} \gamma^{5}\tau \psi)^{2}\right] \ .\label{eq6}
\end{aligned}
\end{equation}
Here $N_c$ is the number of color which is given by $N_c=3$ and we only consider the contribution of color singlet terms for simplicity.

From the comparison of Eq.~(\ref{eq2}) [Eq.~(\ref{eq3})] to Eq.~(\ref{eq5}) [Eq.~(\ref{eq6})],
one can see that, with the help of the Fierz transformation, all exchange interaction channels of the original Lagrangian are released. 
In Eq.~(\ref{eq5}), the Fierz transformed Lagrangian contains not only the scalar and pseudoscalar interactions, but also vector and axialvector interaction channels.

Because the Fierz transformation is just a mathematical technique, we can combine the original Lagrangian and Fierz transformed Lagrangian, using a weighting factor $\alpha$, at any proportion. The factor $\alpha$ reflects the competition between the original interaction channels and the exchange interaction channels.
Then the effective Lagrangian becomes
\begin{equation}
\mathcal{L}_{\rm eff}^{~2f}=\bar{\psi}(i \gamma^{\mu}\partial_{\mu} -m+\mu\gamma^{0}) \psi +(1-\alpha)\mathcal{L}_{\mathrm{int}}^{~2f}+\alpha \mathcal{F}(\mathcal{L}_{\mathrm{int}}^{~2f}) \ .\label{eq7}
\end{equation}

Under the mean-field approximation, the mass gap equation and the effective chemical potential can be obtained as follows:
\begin{equation}
\begin{aligned}
M=&m -2\left[(1-\alpha)G +\frac{\alpha G}{12} +\frac{\alpha G_V}{6}\right] \sum_{f=u, d}\sigma_f \\
 =&m -2\left[(1-\alpha) +\frac{\alpha }{12} +\frac{\alpha R_V}{6}\right]G \sum_{f=u, d}\sigma_f \\
 =&m-2 G' \sum_{f=u, d}\sigma_f \ ,
\end{aligned} \label{eq8}
\end{equation}
\begin{equation}
\begin{aligned}
\mu^{*} &=\mu-\left[2(1-\alpha)R_V +\frac{\alpha}{3} +\frac{\alpha R_V}{6} \right] G\sum_{f=u, d}\rho_f \\
 &=\mu-\frac{12R_V +2\alpha-11\alpha R_V}{6} G \sum_{f=u, d}\rho_f \ ,
\end{aligned} \label{eq9}
\end{equation}
where $R_V=G_{V}/G$, $G'=(12-11\alpha+2\alpha R_V)G/12$. The quark condensate $\langle\bar{\psi} \psi\rangle$ and quark number density $\left\langle\psi^{+} \psi\right\rangle$ are denoted as $\sigma$ and $\rho$, respectively, which are the average values of operaters, $\bar{\psi} \psi$ and $\psi^{+} \psi$, in the ground state. 

Eq.~(\ref{eq8}) displays the mechanism of spontaneous chiral symmetry breaking in the NJL model, through which quarks acquire a dynamical mass proportional to the chiral condensates, plus a small contribution due to the bare quark mass. 
Eq.~(\ref{eq9}) demonstrates the effects of vector interactions that quarks obtain an effective chemical potential $\mu^{*}$ which is shifted to a lower value than the physical chemical potential $\mu$.
From Eq.~(\ref{eq8}) and Eq.~(\ref{eq9}), it is clear that the introduction of Fierz transformed identity contributes to the chemical potential and the dynamical quark mass, because the scalar and vector interactions in the exchange channels in Eq.~(\ref{eq5}) and Eq.~(\ref{eq6}) under mean-field approximation is nonzero at finite chemical potential.

\subsubsection{At (zero-temperature) zero chemical potential}

In the present section, we focus on how to obtain the quark condensate and the quark number density, as well as the regularization procedure we used.

At zero temperature and zero chemical potential, the quark condensate has the following form:
\begin{equation}
\begin{aligned}
\sigma_f=\langle\bar{\psi} \psi\rangle_{f} =-\int \frac{\mathrm{d}^{4} p}{(2 \pi)^{4}} \operatorname{Tr}\left[i S_{f}\left(p^{2}\right)\right] \ , \label{eq10}
\end{aligned}
\end{equation}
where the trace ``Tr'' is taken in Dirac and color spaces and the quark propagator of flavor $f$ is
\begin{equation}
\begin{aligned}
S_{f}(p^2) =\frac{1}{\gamma^{\mu} p_{\mu} - M_{f}} \ .\label{eq11}
\end{aligned}
\end{equation}
Note that this is the key equation for the present model because it determines the values of the chiral condensate $\langle\bar{\psi} \psi\rangle_{f}$  and the constituent quark mass $M_{f}$.
Then from evaluating the trace we can obtain:
\begin{equation}
\begin{aligned}
\sigma_{f} =- N_{\mathrm{c}} \int_{-\infty}^{+\infty} \frac{\mathrm{d}^{4} p}{(2 \pi)^{4}} \frac{4 i M_{f}}{p^{2}-M_{f}^{2}} \ . \label{eq12}
\end{aligned}
\end{equation}
We mention here that the previous calculations are usually performed in Minkowski space.
To perform the present calculations, we employ a Wick rotation from Minkowski space to Euclidean space
and find correspondingly:
\begin{equation}
\sigma_{f}=-N_{\mathrm{c}} \int_{-\infty}^{+\infty} \frac{\mathrm{d}^{4} p^{\mathrm{E}}}{(2 \pi)^{4}} \frac{4  M_{f}}{\left(p^{\mathrm{E}}\right)^{2}+M_{f}^{2}}. \label{eq13}
\end{equation}

Because of the fact that we simplified the interactions as four-fermion contact point-like interactions (or six-fermion interactions as well in the case of $uds$ quark matter) in the Lagrangian, the NJL model cannot be renormalized, and the condensate will be divergent, as one can observe in Eq.~(\ref{eq13}). 
Consequently, it must be interpreted as an effective field theory, which is only valid up to a certain cutoff energy scale $\Lambda$. 
For the purpose of the present study, the energy cutoff should be far beyond the baryon chemical potential possibly reached in some massive quark stars $\Lambda \gg \mu_{\rm B}$, with $\mu_{\rm B}\lesssim 1.5$ GeV from our calculations (see below).
The parameter $\Lambda$ can also be interpreted as the scale at which the strong interaction vanishes, a crude approximation for the property of asymptotic freedom of QCD. 
Then, to avoid the ultraviolet (UV) divergence and make the integral finite, a certain regularization scheme is inevitable. We note that this is an acceptable procedure when the chemical potential $\mu_q$ we considered is less than the cutoff $\Lambda$ so as to get a reliable result. 
There are several regularization procedures that can be used to deal with the UV divergence, such as the three momentum cutoff in 3-momentum space, which is mathematically convenient. Still, this method is a sharp cutoff and has the disadvantage of being not covariant. 
Here, we adopt the proper-time regularization (PTR) with a UV cutoff, which can not only allow the momentum integral up to infinity but also avoid the UV divergence with a \textit{soft} cutoff. 
Furthermore, PTR also has the features that it is invariant and has an $O(3)$ symmetry for $\mu \neq 0$, while for the limit $\mu \rightarrow 0$, the $O(4)$ symmetry is restored.

By definition, the PTR is equal to replace the ultraviolet divergent integrand $1/A_{n}$ as an integral of its exponential function, that is
\begin{equation}
\begin{aligned}
&\frac{1}{A^{n}}=\frac{1}{(n-1) !} \int_{0}^{\infty} \mathrm{d} \tau \tau^{n-1} e^{-\tau A} \\
&\stackrel{\text { UV cutoff }}{\longrightarrow} \frac{1}{(n-1) !} \int_{\tau_{\mathrm{UV}}}^{\infty} \mathrm{d} \tau \tau^{n-1} e^{-\tau A} \ ,  \label{eq14}
\end{aligned}
\end{equation}
where $\Lambda_{\rm UV}$ is the parameter related to ultraviolet cutoff. The lower cutoff $\tau_{\rm UV}=1/\Lambda_{\rm UV}^2$ induces the dumping factor into the original propagator, therefore high frequency contribution is dumped, so the original divergent integral turns out to be finite.
Then, we obtain:
\begin{equation}
\begin{aligned}
\sigma_{f}
&=-N_{\mathrm{c}} \int_{-\infty}^{+\infty} \frac{\mathrm{d}^{4} p^{\mathrm{E}}}{(2 \pi)^{4}} \frac{4  M_{f}}{\left(p^{\mathrm{E}}\right)^{2}+M_{f}^{2}} \\
&=-\frac{3 M_{f}}{4 \pi^{2}} \int_{\tau_{\mathrm{UV}}}^{\infty} \mathrm{d} \tau \frac{e^{-\tau M_{f}^{2}}}{\tau^{2}} \ . \label{eq15}
\end{aligned}
\end{equation}

\subsubsection{At (zero-temperature) finite chemical potential}

In Euclidean space, introducing the chemical potential at zero temperature is equivalent to perform a transformation~\cite{2005PhRvC..71a5205Z,2005PhR...407..205B}:
$p_{4} \rightarrow p_{4}+i \mu_{f}^{*}$.
Then, after integrating over $p_{4}$ first and applying proper-time regularization, one can obtain the analytical results of quark condensate as follows:
\begin{widetext}
\begin{equation}
\begin{aligned}
&\sigma_{f}=
-N_{\mathrm{c}} \int_{-\infty}^{+\infty} \frac{\mathrm{d}^{4} p^{\mathrm{E}}}{(2 \pi)^{4}} \frac{4  M_{f}}{\left(p^{\mathrm{E}}\right)^{2}+M_{f}^{2}}
=-N_{\mathrm{c}} \int_{-\infty}^{+\infty} \frac{\mathrm{d}^{4} p^{\mathrm{E}}}{(2 \pi)^{4}} \frac{4  M_{f}}{\left(  p_{4}+ i\mu_{f}^{*} \right)^{2}+ p^{2}+M_{f}^{2}}
\\
&
=-\frac{3 M_{f}}{\pi^{3}} \int_{0}^{+\infty} \mathrm{d} p \int_{-\infty}^{+\infty} \mathrm{d} p_{4} \frac{p^{2}}{\left(p_{4}+i \mu_{f}^{*} \right)^{2}+M_{f}^{2}+p^{2}}  \\
&= \begin{cases}-\frac{3 M_{f}}{\pi^{2}} \int_{\sqrt{\mu_{f}^{* 2}-M_{f}^{2}}}^{+\infty} \mathrm{d} p \frac{\left[1-\operatorname{Erf}\left(\sqrt{M_{f}^{2}+p^{2}} \sqrt{\tau_{\mathrm{UV}}}\right)\right] p^{2}}{\sqrt{M_{f}^{2}+p^{2}}}, & M_{f}<\mu_{f}^{*} \\ \frac{3 M_{f}}{4 \pi^{2}}\left[M_{f}^{2} \Gamma\left(0, M_{f}^{2}\tau_{\mathrm{UV}} \right) -\frac{e^{-M_{f}^{2} \tau_{\mathrm{Uv}}}}{\tau_{\mathrm{UV}}}\right], & M_{f}>\mu_{f}^{*}\end{cases} \label{eq16}
\end{aligned}
\end{equation}
\end{widetext}
where $\Gamma\left(a, z \right)= \int_{z}^{+\infty} \mathrm{d} t \; t^{a-1} e^{-t}$ and $\operatorname{Erf}(x)=\frac{2}{\sqrt{\pi}} \int_{0}^{x} \mathrm{d} t \; e^{-t^{2}}$.

At zero temperature and finite chemical potential, the quark number density is
\begin{equation}
\begin{aligned}
\rho_{f} &=\left\langle\psi^{+} \psi\right\rangle_{f} =-\int \frac{\mathrm{d}^{4} p}{(2 \pi)^{4}} \operatorname{Tr}\left[i S_{f}\left(p^{2}\right) \gamma_{0}\right] \\
&=2 N_{\mathrm{c}} \int \frac{\mathrm{d}^{3} p}{(2 \pi)^{3}} \theta\left(\mu_{f}^{*}-\sqrt{p^{2}+M_{f}^{2}}\right) \\
&= \begin{cases}\frac{1}{\pi^{2}}\left(\sqrt{\mu_{f}^{* 2}-M_{f}^{2}}\right)^{3} \ , & \mu_{f}^{*}>M_{f} \\
0\ . & \mu_{f}^{*}<M_{f}\end{cases}\ , \label{eq17}
\end{aligned}
\end{equation}

Since there is a step function on the right-hand side of Eq.~(\ref{eq17}), it is obvious that the quark number density of flavor $f$ will vanish when the effective quark chemical potential $\mu^{*}_f$ is smaller than a threshold value (see later in Fig.~\ref{fig:rho}).

\subsection{(2+1)-flavor NJL model} 

For (2+1)-flavor NJL model, the Lagrangian is
\begin{equation}
\mathcal{L}_{\mathrm{NJL}}^{~3f} =\mathcal{L}_{0}+\mathcal{L}_{\mathrm{int}}^{~3f} \ , \label{eq18}
\end{equation}
where the interaction term is written as: 
$\mathcal{L}_{\mathrm{int}}^{~3f} =\mathcal{L}_{\sigma}^{4} +\mathcal{L}_{\sigma}^{6}+ \mathcal{L}_{V}^{4}$. 
The phenomenological vector interaction term $\mathcal{L}_{V}^{4}$ is the same as in Eq.~(\ref{eq3})
and 
\begin{equation}
\mathcal{L}_{\sigma}^{4} =\sum_{i=0}^{8} G\left[\left(\bar{\psi} \lambda_{i} \psi\right)^{2}+\left(\bar{\psi} i\gamma^{5}\lambda_{i} \psi\right)^{2}\right] \ . \label{eq19}
\end{equation}
 The six-fermion interaction term is written as: 
\begin{equation}
\mathcal{L}_{\sigma}^{6} = -K\left(\operatorname{det}\left[\bar{\psi}\left(1+\gamma^{5}\right) \psi\right]+\operatorname{det}\left[\bar{\psi}\left(1-\gamma^{5}\right) \psi\right]\right) \ . \label{eq20}
\end{equation}
It represents the effects of the instanton-induced QCD axial anomaly, which is a determinant in flavor space and breaks the $U(1)_{A}$ axial symmetry of the QCD Lagrangian.
$G$ and $K$ are the four-fermion and six-fermion interaction coupling constants, respectively. $\lambda_{i}\; (i=1 \rightarrow 8)$ is the Gell-Mann matrix in flavor space. 
$\lambda_{0}= \sqrt{2/3}\; I_{0}$ ($I_{0}$ is the identity matrix).

The Fierz identity of the interaction terms in the (2+1)-flavor NJL model is
\begin{equation}
\mathcal{F}(\mathcal{L}_{\mathrm{int}}^{3f} )=\mathcal{F}(\mathcal{L}_{\sigma}^{4})+ \mathcal{F}(\mathcal{L}_{\sigma}^{6})+\mathcal{F}(\mathcal{L}_{V}^{4}) \ .\label{eq21}
\end{equation}
Applying the Fierz transformation to four-fermion scalar and pseudoscalar interaction term $\mathcal{F}(\mathcal{L}_{\sigma}^{4})$ and only considering the contributions of color-singlet terms, the Fierz identity can be obtained as follows:
\begin{equation}
\mathcal{F}(\mathcal{L}_{\sigma}^{4})=-\frac{3 G}{2 N_{c}}\left[\left(\bar{\psi} \gamma_{\mu} \lambda_{i}^{0} \psi\right)^{2}-\left(\bar{\psi} \gamma_{\mu} \gamma_{5} \lambda_{i}^{0} \psi\right)^{2}\right] \ .\label{eq22}
\end{equation}
The Fierz identity of the phenomenological vector interaction $\mathcal{F}(\mathcal{L}_{V}^{4})$ is the same as Eq.(~\ref{eq6}).
Note that the six-fermion interaction term does not change after the Fierz transformation, because that Fierz transformation of six-fermion interaction can be defined as transformation that leaves the interaction invariant under all possible permutations of the quark spinors $\psi$ occurring in it~\cite{1992RvMP...64..649K}. That is to say:
\begin{equation}
\mathcal{F}(\mathcal{L}_{\sigma}^{6})=\mathcal{L}_{\sigma}^{6} \ .\label{eq23}
\end{equation}

Then the effective Lagrangian becomes:
\begin{equation}
\mathcal{L}_{\rm eff}^{~3f}=\bar{\psi}(i \gamma^{\mu}\partial_{\mu} -m+\mu\gamma^{0}) \psi +(1-\alpha)\mathcal{L}_{\mathrm{int}}^{~3f}+\alpha \mathcal{F}(\mathcal{L}_{\mathrm{int}}^{~3f}) \ .\label{eq24}
\end{equation}

Under the mean-field approximation, we can obtain the mass gap equations and the effective chemical potential $\mu_{f}^{*}$ of flavor $f$ as follows: 
\begin{equation}
\begin{aligned}
M_{f}&= m_{f}-4 \left[(1-\alpha) +\frac{1}{6}\alpha R_V\right]G \sigma_{f}+2 K \sigma_{j} \sigma_{k}\\
&= m_{f}-4G'\sigma_{f}+2 K \sigma_{j} \sigma_{k} , \label{eq25}
\end{aligned}
\end{equation}
\begin{equation}
\begin{aligned}
\mu_{f}^{*}&=\mu_{f} -\left[2(1-\alpha)G_{V}+\frac{2}{3}\alpha G\right]\sum_{f^{\prime}=u, d, s}\rho_{f^{\prime}}-\frac{1}{3}\alpha G_{V}\rho_{f}\\
&=\mu_{f} -\left[2(1-\alpha)R_V +\frac{2}{3}\alpha\right]G\sum_{f^{\prime}=u, d, s}\rho_{f^{\prime}} -\frac{1}{3}\alpha R_V G \rho_{f} \ .\label{eq26}
\end{aligned}
\end{equation}
where we define $G'=(6-6\alpha+\alpha R_V)G/6$ and $f, j, k $ are the even permutations of $u, d, s$.
At finite chemical potential and zero temperature, the expressions of the quark condensate $\sigma_{f}$ and quark number density $\rho_{f}$ are the same as Eq.~(\ref{eq16}) and Eq.~(\ref{eq17}), respectively.

\begin{table}
\centering
\caption{NJL model parameters satisfying the constraints on the current quark masses $m_u$ and $m_s$ from the recent Review of Particle Physics~\cite{2020PTEP.2020h3C01P}. The units of the coupling constants $G'$ and $K$ are $\mathrm{MeV}^{-2}$ and $\mathrm{MeV}^{-5}$, respectively, and the other parameters 
have the units of $\mathrm{MeV}$.}
         \vskip+2mm
\renewcommand\arraystretch{1.5}
\begin{ruledtabular}
\begin{tabular*}{\hsize}{@{}@{\extracolsep{\fill}}lccccc@{}}
 & $m_{u}$ & $m_{s}$ & $\Lambda_{\mathrm{UV}}$ & $G'$ & $K$   \\
\hline Two flavor  & $3.3$ & /\multirow{2}{*} & $1330$ & $2.028 \times 10^{-6}$ &
/ \\
\hline $2+1$ flavor & $3.4$ & 104 & 1330 & $1.51 \times 10^{-6}$ & $2.75 \times 10^{-14}$ \\
\end{tabular*}
\end{ruledtabular}
    \vspace{-0.4cm}
\label{table:1}
\end{table}

\subsection{Parameter fixing in the NJL models}

From Eqs.~(\ref{eq8}-\ref{eq9}) and Eqs.~(\ref{eq25}-\ref{eq26}), it is clear that the introduction of Fierz transformed identity contributes to the renormalized chemical potential and the gap equation. After defining the new coupling constant $G'$, and keeping the expression of gap equation the same as the widely used one in Eq.~(\ref{eq8}) and Eq.~(\ref{eq25}), at a zero temperature and chemical potential, apart from $\alpha$ and $R_V$, the fixing of the model parameters is the same with the original version of the NJL model~\cite{1994PhR...247..221H}.
According to the latest edition of the Review of Particle Physics Ref.~\cite{2020PTEP.2020h3C01P}, the current quark mass $m_u$ and $m_s$ are predicted to be $\bar{m}=\left(m_u+m_d\right) / 2=3.5_{-0.2}^{+0.5} \mev$ and $m_s=95_{-3}^{+9} \mev$ respectively. Similar to the procedure in Ref.~\cite{1994PhR...247..221H}, after fixing the masses of the up and down quarks by equal values, the other parameters $m_s, \Lambda_{\mathrm{UV}}, G', K$ are chosen to reproduce the experimental data of the pion decay constant and pion mass for $f_{\pi}=92\mev,\,M_{\pi}=135\mev,\,M_{K^{0}}=495\mev,\,M_{\eta}=548\mev,\, M_{\eta^{\prime}}=958\mev$. 
\begin{figure}
\centering
\includegraphics[width=0.49\textwidth]{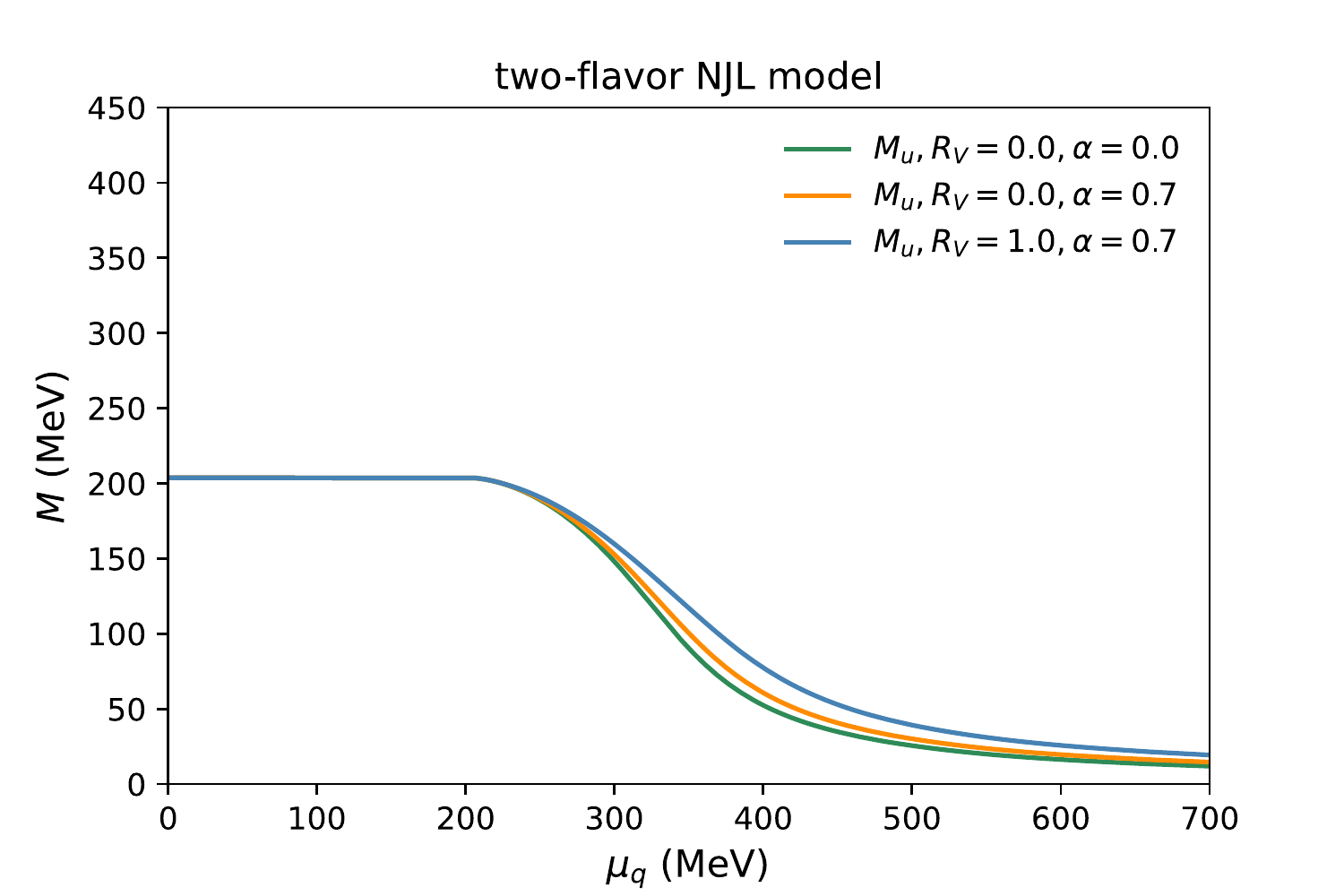}
\includegraphics[width=0.49\textwidth]{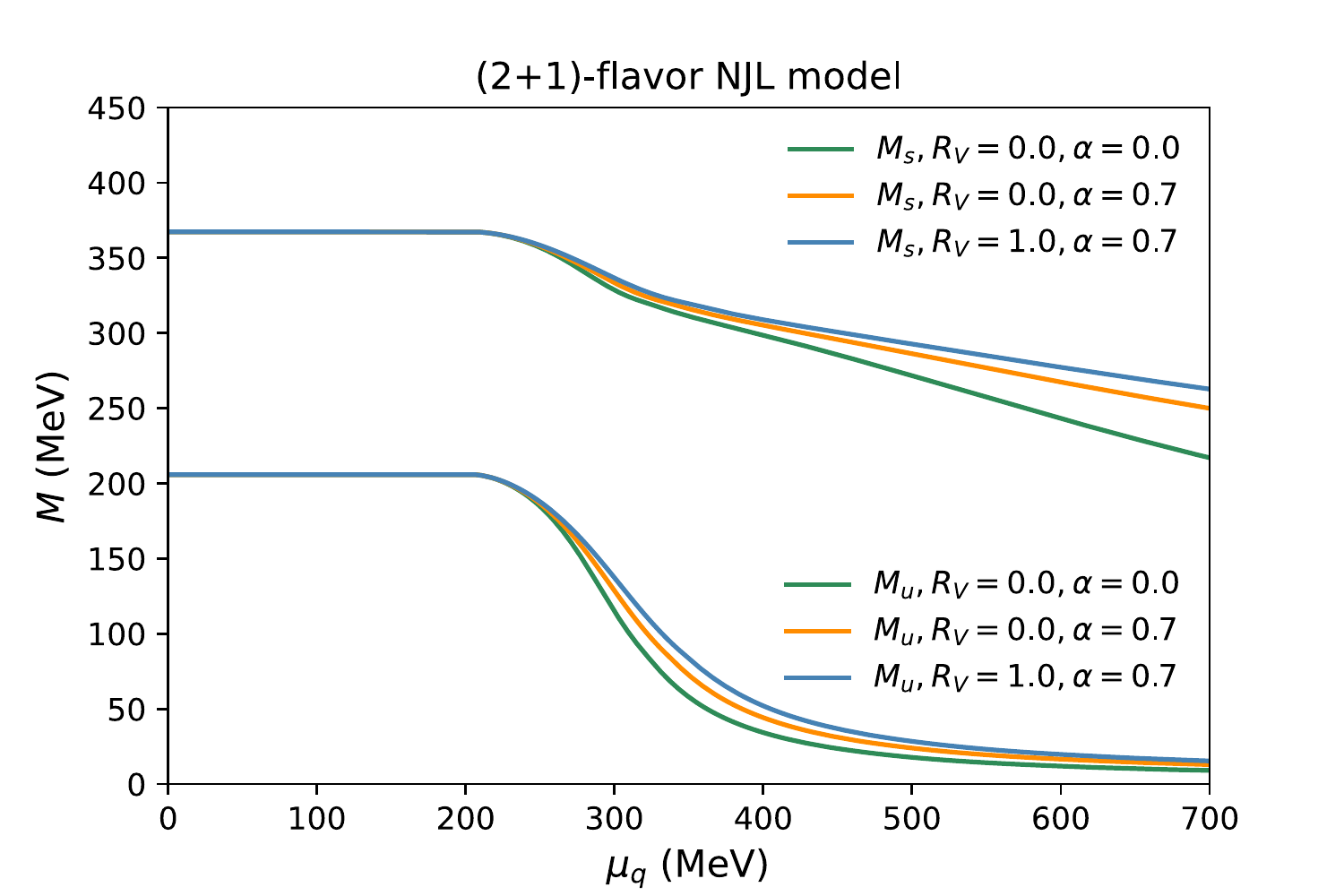}
    \vskip-2mm
\caption{Constituent quark mass of $u$, $d$ and $s$ quark versus quark chemical potential $\mu_q$ for two-flavor and (2+1)-flavor NJL models. 
The results for three representative sets of ($R_V, \alpha$) parameters are shown: ($R_V=0.0, \alpha=0$), ($R_V=0.0, \alpha=0.7$), ($R_V=1.0, \alpha=0.7$).
}\label{fig:mass}
    \vspace{-0.4cm}
\end{figure}

The description still contains two undetermined parameters: $\alpha$ and $R_{V}$.
The free parameter $\alpha$, important to the determination of the chemical potential and the dynamical quark mass, is found to affect 
the EoS of quark matter and can be possibly constrained from the stellar properties composed of the matter.
As for the parameter $R_{V}$, there are many uncertainties. In Ref.~\cite{2012PhRvD..85k4017S}, a ratio around $R_V=0.2$ between the vector and scalar coupling was obtained from an evaluation of only the Fock contributions of the scalar channels. Values in the range $0.25<R_{V}<0.5$ were derived by a Fierz transformation of effective one-gluon exchange interaction, with $G_{V}$ depending on the strength of the $U_{A}(1)$ anomaly in the two-flavor model~\cite{1992RvMP...64..649K,2011PhRvD..84e6010K}. Other attempts to estimate $G_{V}$ are based on the fitting of the vector meson spectrum~\cite{1991PrPNP..27..195V}. However, the relation between the vector coupling in dense quark matter and the meson spectrum in vacuum is expected to be strongly modified by in-medium effects (see discussions in Refs.~\cite{2008PhRvD..77k4028F,2008PhRvD..78c9902F,2009PhRvD..80a4015Z}). 
As a result, presently, the coupling strength of the direct term cannot be fixed, so the total effects of the vector interactions are still unknown.
Because of the uncertainties discussed above, in the present study, we treat both $\alpha$ and $R_{V}$ as free parameters and aim to seek information on them from the astronomical observations of compact stars (see below in Sec. \ref{sec:star}).

\begin{figure}
\centering
\includegraphics[width=0.49\textwidth]{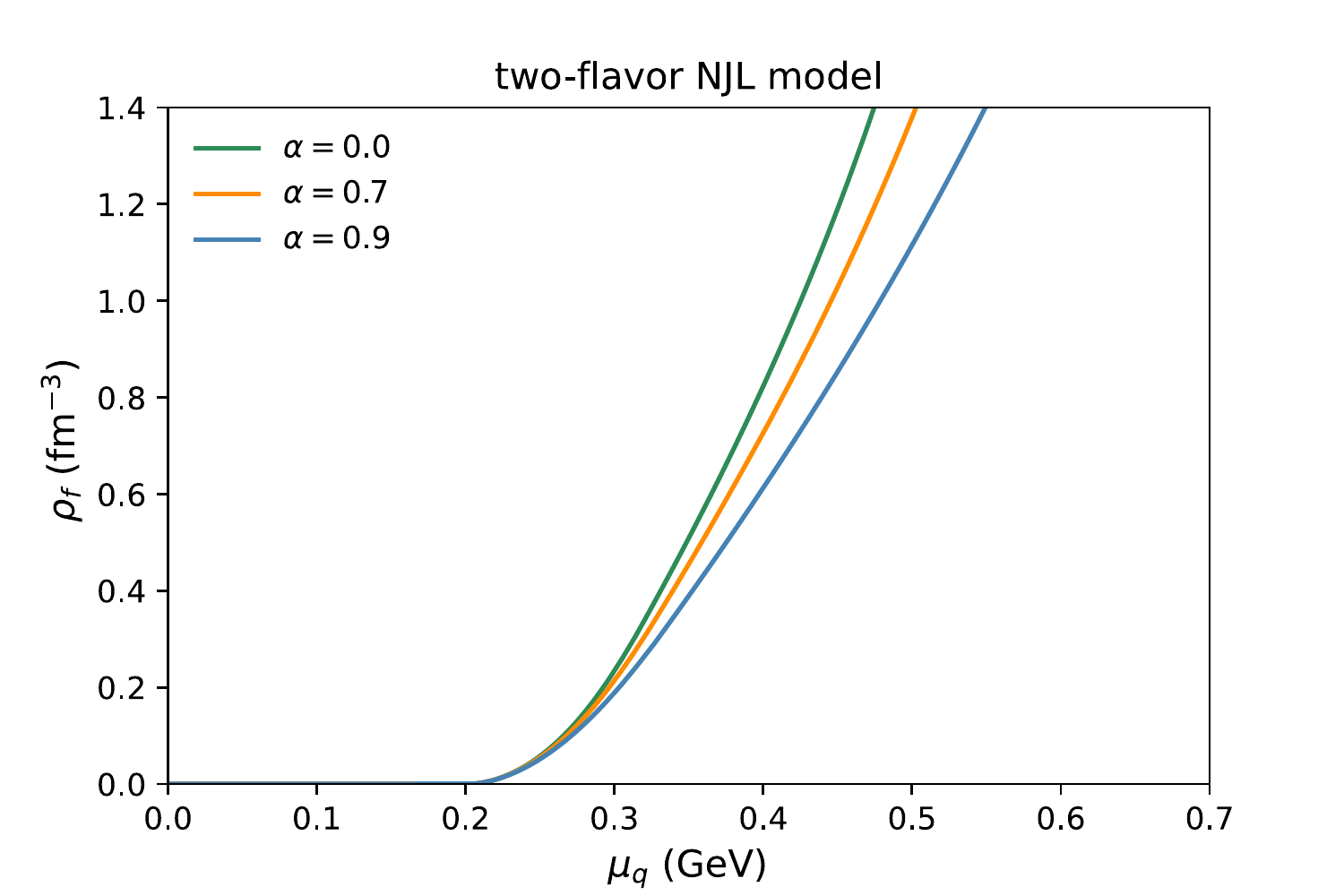}
\includegraphics[width=0.49\textwidth]{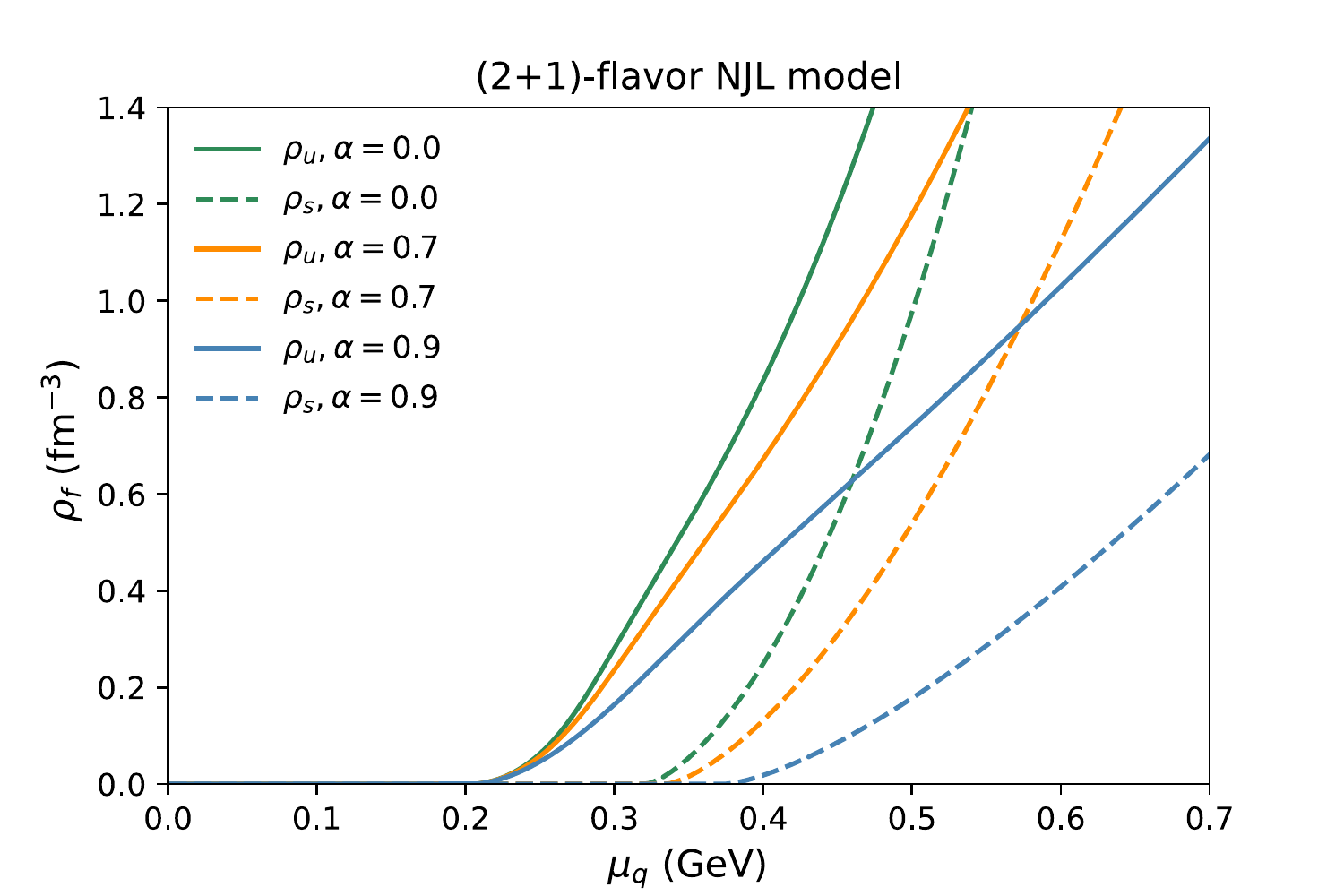}
         \vskip-2mm
\caption{Quark number density $\rho_f$ ($f=u,d,s$) versus quark chemical potential within both two-flavor and (2+1)-flavor NJL models. The calculations are done for representative cases of $\alpha=0.0,~0.7,~0.9$ at $R_V=0.0$.
}    \vspace{-0.4cm}
\label{fig:rho}
\end{figure}
The employed model parameters are collected in Table~\ref{table:1}. 
From solving the mass gap equations of Eq.~(\ref{eq8}) [Eq.~(\ref{eq25})] for two-flavor and (2+1)-flavor NJL models, we can the obtain the dynamical quark masses as functions of the quark chemical potential, which are reported in Fig.~\ref{fig:mass}.
When $\mu_{f}^{*}<M_{f}$, the quark condensate is independent of the quark chemical potential (see Eq.~(\ref{eq16})), correspondingly there is a plateau in Fig.~\ref{fig:mass} for both two-flavor and (2+1)-flavor cases. 
And there is no strangeness at low chemical potential simply due to the relatively large mass of the strange quarks. 
In addition, as the chemical potential increases, the vector interactions make the dynamical quark mass decrease slightly slowly, in comparison with the standard NJL model without vector interactions ($R_V=0.0,\,\alpha=0.0$). 
 Let us focus on the lower panel of Fig.~\ref{fig:mass} for (2+1)-flavor NJL model. 
When $\mu_{u,d} <207\mev$, or $\mu_s<210\mev$, the constituent quark mass stays the same as in the vacuum case where quarks are strong interacted and confined. 
With the increase of the chemical potential, the constituent masses of $u$ and $d$ quarks decrease more quickly than the $s$ quarks.
In particular, when $\mu<320\mev$, the decrease of the dynamical quark mass for the $s$ quarks is due to a flavor-mixing effect and related to the drop of $M_u$ and $M_d$. 
Above $320\mev$, this contribution can be neglected, and $M_s$ starts to decrease again when the number density of the strange quarks $\rho_s$ becomes nonzero. 
Furthermore, we can see that when $\mu_{u,d}>500\mev$, the constituent mass of the $u$ and $d$ quarks change much more slowly than before and not change at large chemical potential,
and the quark mass restores to its current mass, with quarks weakly interacting and deconfined.
As for the two-flavor case in Fig.~\ref{fig:mass}, 
the constituent quark mass holds its vacuum value when $\mu<210\mev$ and changes much slowly near $\mu>500\mev$.

After deriving the constituent quark masses of the $u$, $d$ and $s$ quarks, 
following Eq.~(\ref{eq17}), one can immediately obtain the quark number density for each flavor of quarks.
To show the contribution of the Fierz identity at finite chemical potential, we set $R_V=0$ in both two-flavor and (2+1)-flavor NJL model Lagrangian, that is to say, only considering the standard NJL model and the contribution from the exchanging channels.
Namely, 
$\mathcal{L}_{\rm eff}^{~2f}=\bar{\psi}(i \gamma^{\mu}\partial_{\mu} -m+\mu\gamma^{0}) \psi +(1-\alpha)\mathcal{L}_{\mathrm{\sigma}}^{~2f}+\alpha \mathcal{F}(\mathcal{L}_{\mathrm{\sigma}}^{~2f})$,  and
$\mathcal{L}_{\rm eff}^{~3f}=\bar{\psi}(i \gamma^{\mu}\partial_{\mu} -m+\mu\gamma^{0}) \psi +(1-\alpha)(\mathcal{L}_{\sigma}^{4} +\mathcal{L}_{\sigma}^{6})+\alpha \mathcal{F}(\mathcal{L}_{\sigma}^{4} +\mathcal{L}_{\sigma}^{6})$.
Then we report the quark number densities as functions of the quark chemical potential in Fig.~\ref{fig:rho}.

As previously expected, 
the quark number density stays zero when $\mu_q$ is smaller than the constituent quark mass $M_{u,d}$ and $M_s$.  
Once $\mu_q$ is above some threshold $\mu_c$, the quark number density becomes a monotonically increasing function of $\mu_q$. 
Let us again focus on the results of the (2+1)-flavor NJL model in the lower panel of Fig.~\ref{fig:rho}. One can see that the threshold for which the quark number densities turn to be nonzero for the $u$ and $d$ quarks is around $205\mev$;
The threshold for the $s$ quarks is larger due to their larger vacuum mass.
When we increase the contribution of the exchange channels, 
namely increasing $\alpha$, the threshold for which the quark number density stars to appear is pushed to even higher chemical potential, due to the strong vector repulsion at large $\alpha$. 
%
\section{Quark matter and quark stars} \label{sec:star}

In this section, we compute the EoS of quark matter, as well as the global properties of self-bound strange and nonstrange quark stars, based on the two-flavor and (2+1)-flavor NJL models described in Sec. \ref{sec:form}.

\subsection{QCD vacuum pressure and the bag constant}    \label{sec:bag}
    
The study of the partition function is at the crux of equilibrium statistical field theory. The thermodynamic properties of a system, such as the EoS, are completely determined by the partition function. 
At finite chemical potential and zero temperature, the pressure-versus-chemical-potential relation for quark matter can be strictly proved with the functional path integrals of QCD~\cite{2008PhRvD..78e4001Z,2008IJMPA..23.3591Z} as the expression shown below:
\begin{equation}
P(\mu; M)=P(\mu=0; M)+\int_{0}^{\mu} d \mu^{\prime} \rho\left(\mu^{\prime}\right) . \label{eq27}
\end{equation}
Here the first term $P(\mu=0; M)$ is the pressure at $\mu=0$, which is density-independent quantity and represents the vacuum pressure.
$M$ is a solution of the gap equation shown before. The second term contains all the nontrivial $\mu$-dependence. Note that the formula of Eq.~(\ref{eq27}) is formally model-independence. 
At present, it is difficult to calculate the $P(\mu=0; M)$ from the first-principles QCD, 
therefore when applying Eq.~(\ref{eq27}) to calculate the EoS of the QCD matter, one has to
make use of various non-perturbative QCD models.

\begin{figure*}
\centering
{\includegraphics[width=0.48\textwidth]{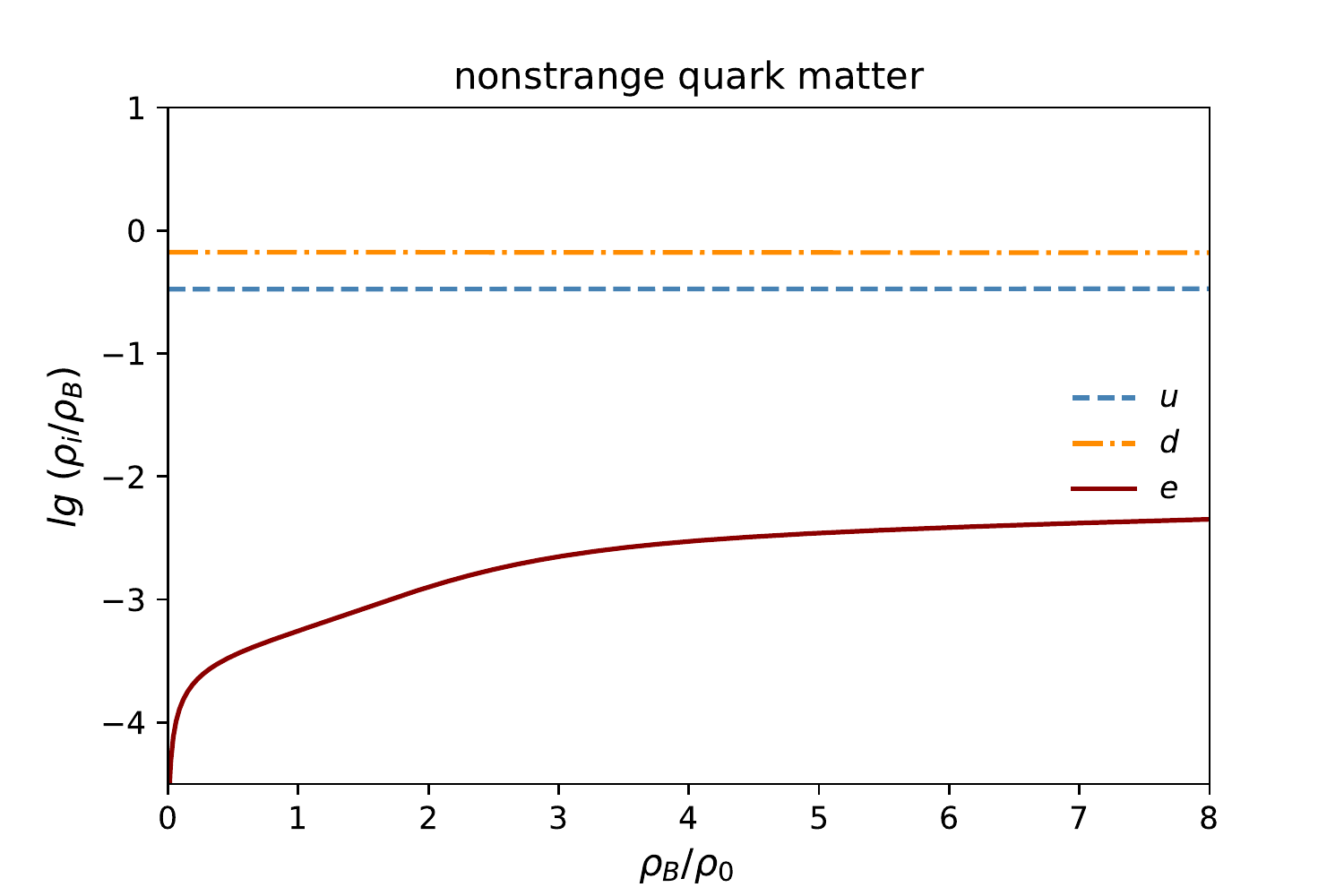}}
{\includegraphics[width=0.48\textwidth]{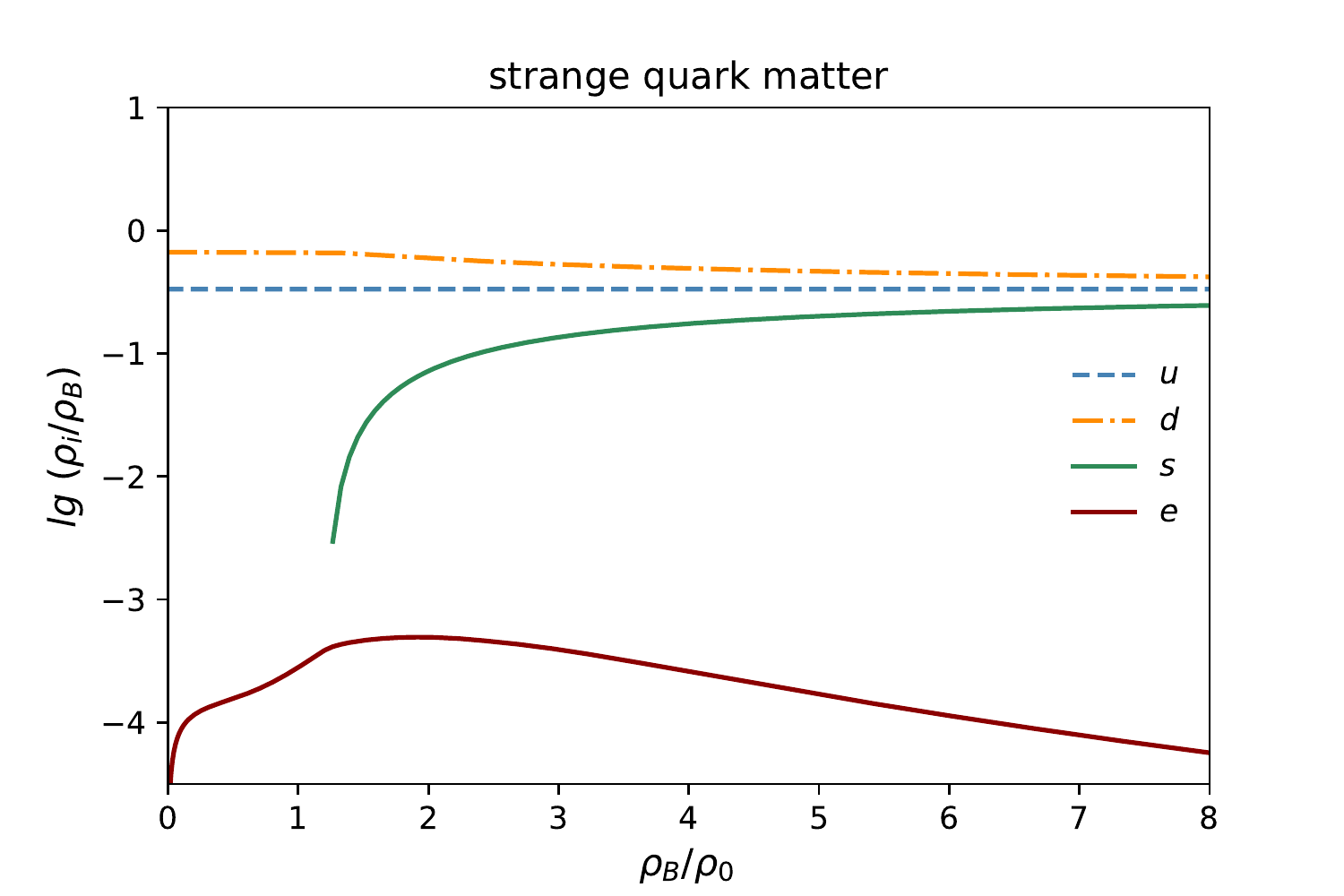}}
         \vskip-2mm
\caption{Logarithmic relative fractions of different constituents $\rho_i/\rho_{\rm B}$ ($i=u, d, s, e$) in chemically equilibrated, charge neutral nonstrange (left panel) and strange (right panel) quark matter, calculated using the two-flavor NJL model ($R_V=0, \alpha=0.82, B^{1/4}=106.6 \mev$) and (2+1)-flavor NJL model ($R_V=0, \alpha=0, B^{1/4}=111 \mev$), respectively. $\rho_0=0.16~\fm3$ is the nuclear saturation density.}\label{fig:fraction}
    \vspace{-0.4cm}
\end{figure*}

Since the vacuum pressure $P(\mu=0; M)$ is not a measurable quantity, one can only evaluate the vacuum pressure difference with respect to a reference ground state. The reference ground state should, in principle, be a trivial vacuum of the strong interaction system that we are studying.
In the NJL-type models, people usually denote the trivial vacuum as $P(\mu; m)$, where $m$ is the current quark mass, and use the parameter $B$ to describe the pressure difference between the trivial and the non-trivial vacuum (Nambu vacuum, reflecting the spontaneous symmetry breaking of the vacuum). Thus the vacuum pressure $B$, as a dynamical consequence of the interaction, can be calculated consistently in the NJL-type models~\cite{2005PhR...407..205B}.

However, such a procedure of determining the bag constant is somewhat unsatisfying, since the pressure computed within the NJL-type models at vanishing density is used in a regime where the model cannot be trusted due to its lack of confinement. Therefore, following the previous studies \cite{2008PhRvD..77f3004P,2012ApJ...759...57L}, we take $P(\mu=0; M)$ as a phenomenological parameter corresponding to $-B$ (vacuum bag constant), which preserves the confinement of quarks. Namely, the bag constant is introduced similarly as in the MIT bag model: $P(\mu=0)=-B$. 
Then, from a known quark number density $\rho(\mu)$ of each flavor, which matches the phenomena of QCD, one can obtain the pressure that satisfies the behavior of QCD at finite chemical potential $\mu$.
Eq.~(\ref{eq27}) tells us that, when $\mu<\mu_c$, the pressure $P$ equals $-B$, thus for $P>0$, the chemical potential starts from a nonzero value. 
It should be stressed that the EoS of the strong-interaction matter depends not only on the Nambu solution but also on the vacuum pressure, reflecting the non-perturbative vacuum nature of QCD. The non-perturbative vacuum plays a vital role in the study of the compact star structures, as shown in many previous studies (see recent discussions in e.g., Refs.~\cite{2021PhRvD.103f3018Z,2021EPJC...81..921L,2021arXiv211209595P}).

\begin{figure*}
\centering
\includegraphics[width=0.49\textwidth]{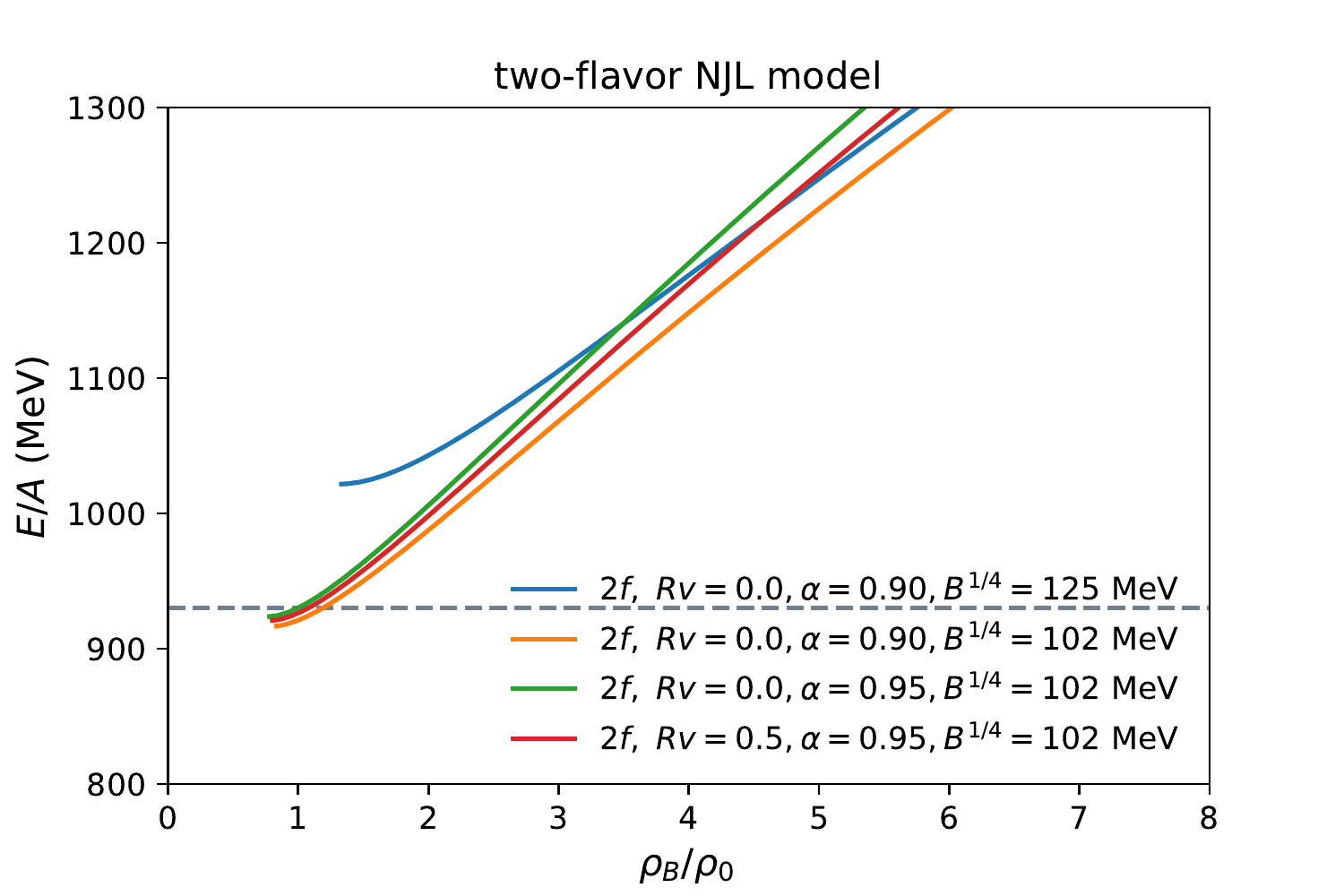}
\includegraphics[width=0.49\textwidth]{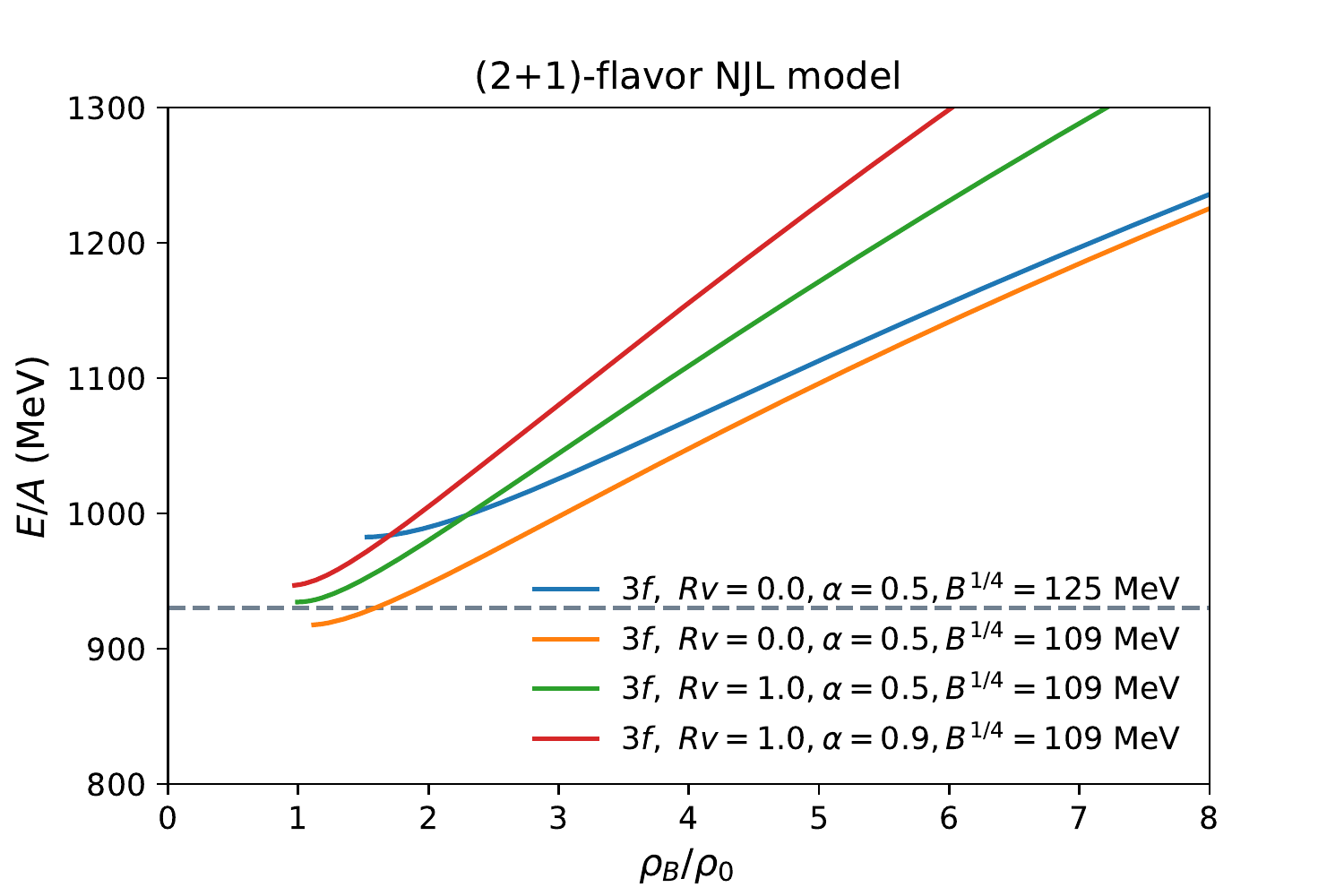}
\caption{Energy per baryon versus density (in units of the nuclear saturation density $\rho_0$) for various choices of the two-flavor and (2+1)-flavor NJL model parameters for the study of their effects (see text for details). The horizontal lines in both panels refer to the energy per baryon of the most stable nuclei known, $E/A (^{56}\rm Fe)=930\mev$.
}\label{fig:binding}
    \vspace{-0.4cm}
\end{figure*}
%
\subsection{Stability of self-bound quark matter and the equation of state}  \label{sec:stability}

Quark matter is in equilibrium with respect to the weak-interaction processes, \vskip-6mm
\begin{eqnarray}
&& d \rightarrow u + e + \tilde{\nu}_e\ ,~u + e \rightarrow d + \nu_e\ ;\nonumber \\
&& s \rightarrow u + e + \tilde{\nu}_e\ ,~u + e \rightarrow s + \nu_e\ ;\nonumber \\
&& s + u \leftrightarrow d + u\ . \nonumber
\end{eqnarray}
The $\beta$-stable conditions, \vskip-4mm
\begin{equation}
\mu_s = \mu_d = \mu_u + \mu_e \ , \label{eq28}
\end{equation} 
should be fulfilled. 
One has to require also the charge-neutrality of the quark matter, 
\begin{equation}
\frac{2}{3}\rho_u-\frac{1}{3}\rho_d-\frac{1}{3}\rho_s-\rho_e=0 \:,  \label{eq29}
\end{equation}
and the baryon number conservation,
\begin{equation}
\frac{1}{3}\left(\rho_u + \rho_d + \rho_s\right) =\rho_{\rm B} \:, \label{eq30}
\end{equation}
is satisfied with $\rho_{\rm B}$ being the baryon number density. 
Since the electrons are highly relativistic, their particle number density 
is simply
$\rho_e={\mu_e^3}/({3\pi^2})$ . 
The baryon chemical potential is $\mu_{\rm B}=\mu_u +2\mu_d$ and $\mu_{\rm B}=\mu_u +\mu_d +\mu_s$, respectively, for $ud$ and $uds$ matter. 
Due to the constraints of $\beta$ equilibrium and charge-neutrality, there is only one independent chemical potential left. Here, we choose $\mu_u$, and the other chemical potentials, namely $\mu_d$, $\mu_s$ and $\mu_e$ can be treated as a function of $\mu_u$. 
See Fig.~\ref{fig:fraction} for typical compositions of $ud$ and $uds$ quark matter. 

The energy density and pressure of the system have the thermodynamic relation of
\begin{equation}
\varepsilon
=-P+\sum_{i=u, d, s, e} \mu_{i} \rho_{i}\left(\mu_{i}\right). \label{eq31}
\end{equation}
The energy per baryon of quark matter are shown in Fig.~\ref{fig:binding} under various NJL model parameters.  
In general, for larger $B$, $\alpha$ as well as $R_V$, the energy per baryon $E/A$ (or $\varepsilon/\rho_{\rm B}$) all becomes larger. 
The crucial importance of the vacuum bag constant $B$ can be seen from the comparison of the following two cases: When keeping other parameters unchanged, the quark matter is stable for $B=(102~\rm{MeV})^4$ in the two-flavor case [$B=(109~\rm{MeV})^4$ in the (2+1)-flavor case], but unstable for a larger value of $B=(125~\rm{MeV})^4$ in both two-flavor and (2+1)-flavor cases. 
The effect of $B$ is straightforward since, by definition, it is the energy excess between the perturbative and the non-perturbative vacuum;
The effects relating to the $\alpha$ and $R_V$ terms can be understood from their repulsive nature at finite chemical potential [see Eq.~(\ref{eq9})]. 
Nevertheless, resulting from the opposite effects of $B$ and $R_V$ terms on the energy, $E/A$ actually becomes smaller with increasing $R_V$ in the two-flavor case when maintaining the absolute binding of the matter at large values of $\alpha$ (see Fig.~\ref{fig:stable} below).

\begin{figure}
\centering 
\includegraphics[width=0.5\textwidth]{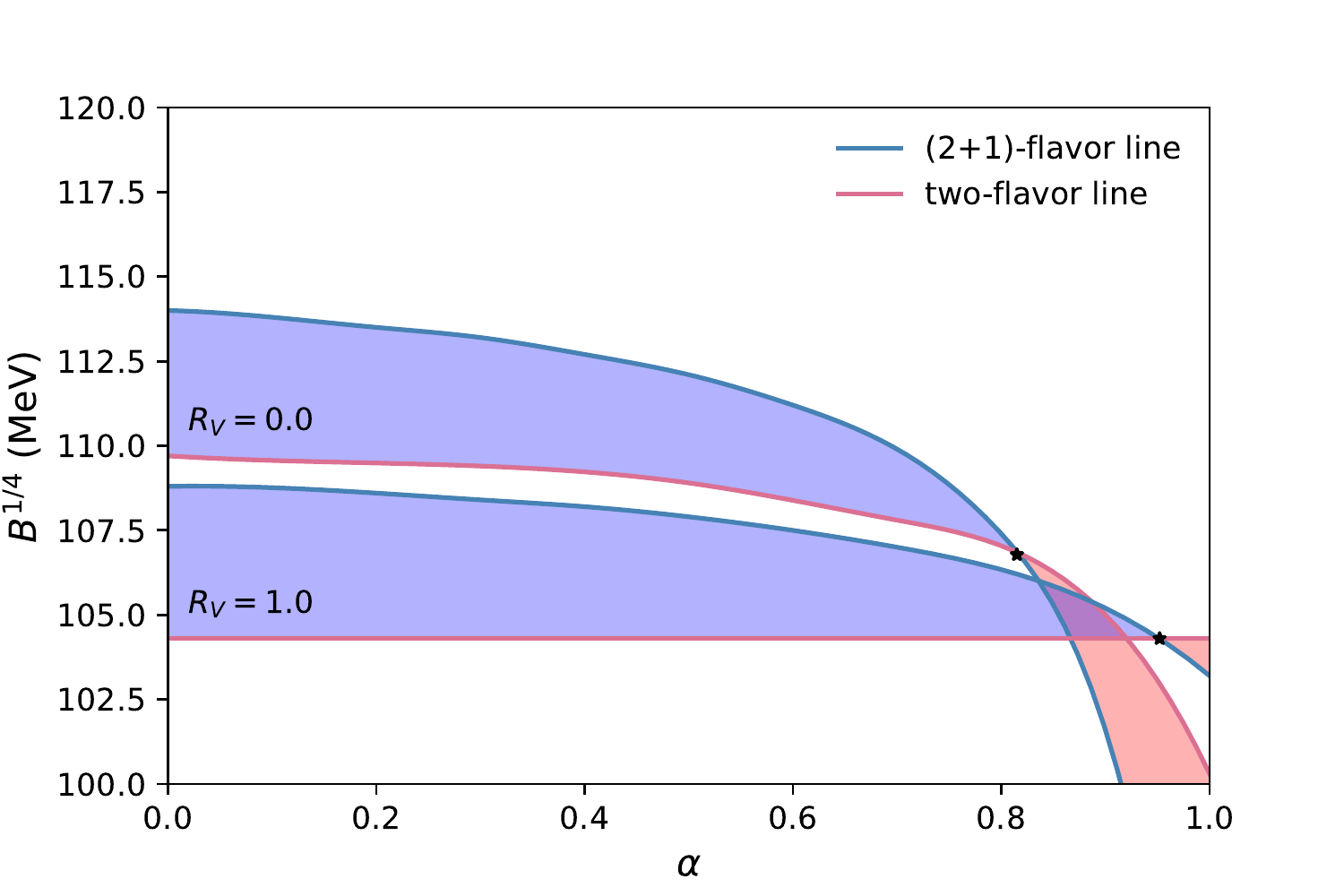}
\caption{Stability windows of $ud$ (pink shaded regions) and $uds$ matter (blue shaded regions) for two cases of vector interaction $R_V=0.0$ and $R_V=1.0$. 
The two-flavor lines in red and (2+1)-flavor lines in blue correspond to $(E/A)_{\rm ud}=930\mev$ and $(E/A)_{\rm uds}=930\mev$, respectively.
The regions above/below each line have a binding energy per particle larger/smaller than $930\mev$, i.e., less/more bound than $^{56}\rm Fe$.
$(0.82, 106.6\mev)$ and $(0.95, 102\mev)$ are the two intersection points for the two-flavor and (2+1)-flavor lines.
    \vspace{-0.4cm}
}\label{fig:stable}
\end{figure}

Since we are interested in studying stellar properties composed of self-bound quark matter, we introduce in detail the stability conditions for limiting our NJL model parameters as follows:
\begin{itemize}
\item For self-bound $uds$ matter,  1) $(E/A)_{\rm uds}\leq930\mev$ to ensure the hypothesis of strange matter to be valid; 2) Atomic nuclei should be stable with respect to the formation of droplets of the $ud$ matter, that is, nonstrange quark matter in bulk should have an energy per baryon higher than that of the confined phase: $(E/A)_{\rm ud}\geq930\mev$; Altogether, $(E/A)_{\rm uds}\leq930\mev\leq(E/A)_{\rm ud}$.
\item For self-bound $ud$ matter, $(E/A)_{\rm ud}\leq930\mev\leq(E/A)_{\rm uds}$.
\end{itemize}
The obtained stability windows of $ud$ and $uds$ matter are shown in Fig.~\ref{fig:stable} in the $B^{1/4}-\alpha$ plane, for two cases of vector interaction: $R_V=0.0$ and $R_V=1.0$. 
In each case, the two-flavor line and (2+1)-flavor line correspond to $(E/A)_{\rm ud}=930\,{\rm MeV}$ and $(E/A)_{\rm uds}=930\,{\rm MeV}$, respectively.
It is evident that, in most cases (small $\alpha$), the two-flavor line is below the corresponding (2+1)-flavor one, which defines the parameter spaces (in blue) for the stable $uds$ matter, satisfying $(E/A)_{\rm uds}\leq930\mev\leq(E/A)_{\rm ud}$.
However, at large $\alpha$, the two-flavor line is actually above the (2+1)-flavor one, which leads to the parameter spaces (in pink) for the stable $ud$ matter, satisfying $(E/A)_{\rm ud}\leq930\mev\leq(E/A)_{\rm uds}$.

\begin{table}
\centering
\caption{Ten employed NJL parameter sets for the study of quark stars in two-flavor and (2+1)-flavor cases, chosen throughout the stability windows of Fig.~\ref{fig:stable}. 
}
\renewcommand\arraystretch{1.5}
\begin{ruledtabular}
\begin{tabular*}{\hsize}{@{}@{\extracolsep{\fill}}lccccc@{}}
$\mathrm{Two~flavor}$ & $1$ & $2$ & $3$ & $4$ & $5$ \\
\hline $R_V$ & $0.0$ & 0.0 & 0.5 & $0.5$ & $0.5$ \\
\hline $\alpha$ & $0.82$ & $1.0$ & $0.95$ & $1.0$ &$1.0$\\
\hline $B^{1/4}$ ($\mathrm{MeV}$)  & $106.6$ & $100.0$ & $102.0$ & $100.0$ &$102.0$\\
\hline \hline $\mathrm{2+1~flavor}$ & $1$ & $2$ & $3$ & $4$ & $5$ \\
\hline $R_V$ & $0.0$ & 0.5 & 0.5 & $0.5$ & $1.0$ \\
\hline $\alpha$ & $0.0$ & $0.0$ & $0.0$ & $0.6$ &$0.95$\\
\hline $B^{1/4}$ ($\mathrm{MeV}$)  & $111.0$ & $107.0$ & $111.0$ & $107.0$ &$104.3$\\ 
\end{tabular*}
\end{ruledtabular}
    \vspace{-0.4cm}
\label{table:2}
\end{table}

\begin{figure}
\centering
\includegraphics[width=0.49\textwidth]{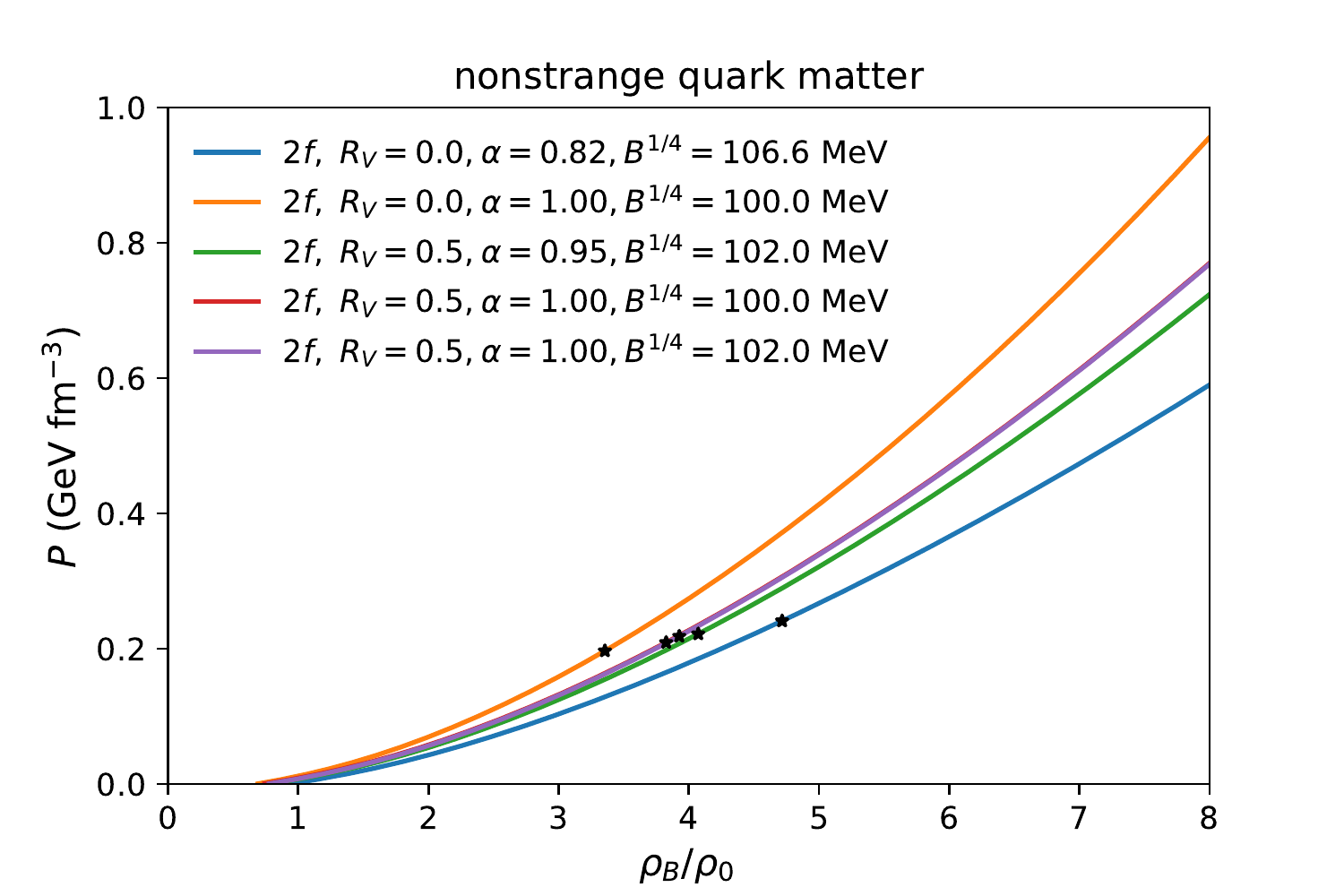}
\includegraphics[width=0.49\textwidth]{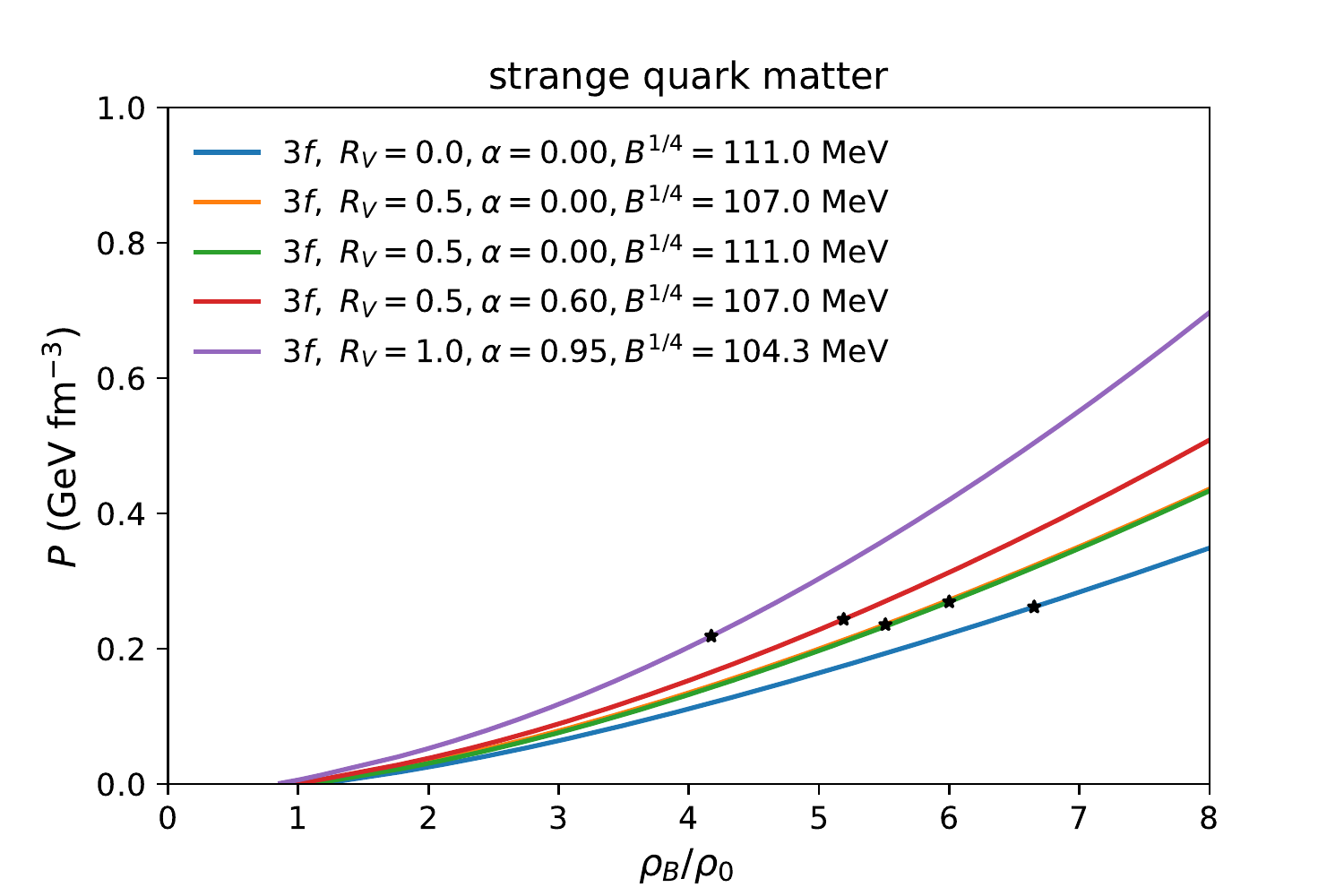}
\caption{Pressure versus density (in units of the nuclear saturation density $\rho_0$) for self-bound quark matter in the present NJL models in both the nonstrange (upper panel) and strange (lower panel) cases. 
For each case, the calculations are done for five representative parameter sets chosen from stability window (see Fig.~\ref{fig:stable}). 
The central densities of the corresponding maximum-mass quark stars are indicated with black stars, respectively.
}    \vspace{-0.4cm}
\label{fig:prho}
\end{figure}

In Fig.~\ref{fig:stable}, the nearly horizontal behaviors of the two-flavor and (2+1)-flavor lines (especially for relatively small $\alpha$) indicate the dominating role of the bag parameter $B$ to the stability of quark matter.
And one observes again the disfavor of the $\alpha$ increase to the binding of $uds$ matter, since the blue-shaded regions shrink with $\alpha$.
This results from the decrease of the $B$ value with the increase of $\alpha$ to maintain the system's binding in both the two-flavor and (2+1)-flavor cases.
Since the $B$ decrease is faster in the (2+1)-flavor case for larger $\alpha$, the $uds$ stability windows finally disappear, and the $ud$ stability windows grow with $\alpha$.
The introduction of the vector interactions shifts down the $uds$ stability windows in the $B^{1/4}$-$\alpha$ plane, due to its repulsive contribution to the energy of the system mentioned above.
Furthermore, it slows the decrease of $B$ with $\alpha$ in both the two-flavor and (2+1)-flavor cases, resulting in a larger $uds$ stability window and a smaller $ud$ stability window. 
For example, the flip point is pushed from $\alpha=0.82$ in the case of $R_V=0$ to $\alpha=0.95$ in the case of $R_V=1.0$.

Hereafter, we employ ten representative NJL parameter sets (five sets in each of the two cases; collected in Table~\ref{table:2}) from the stability windows of Fig.~\ref{fig:stable} to perform the study of strange and nonstrange stellar quark matter and quark stars. 

\begin{figure} 
\resizebox*{0.48\textwidth}{0.25\textheight}
{\includegraphics[width=0.48\textwidth]{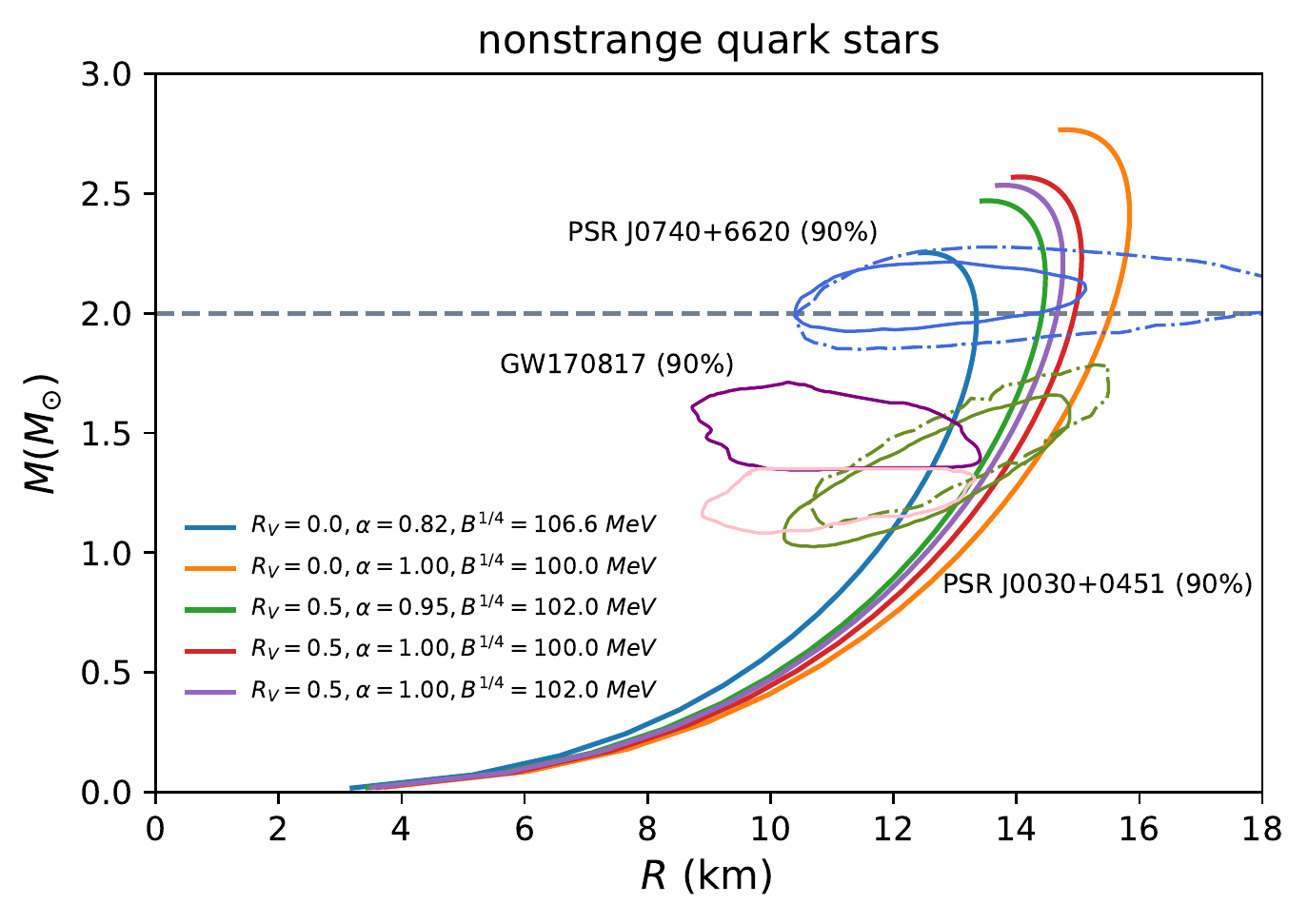}}
{\includegraphics[width=0.48\textwidth]{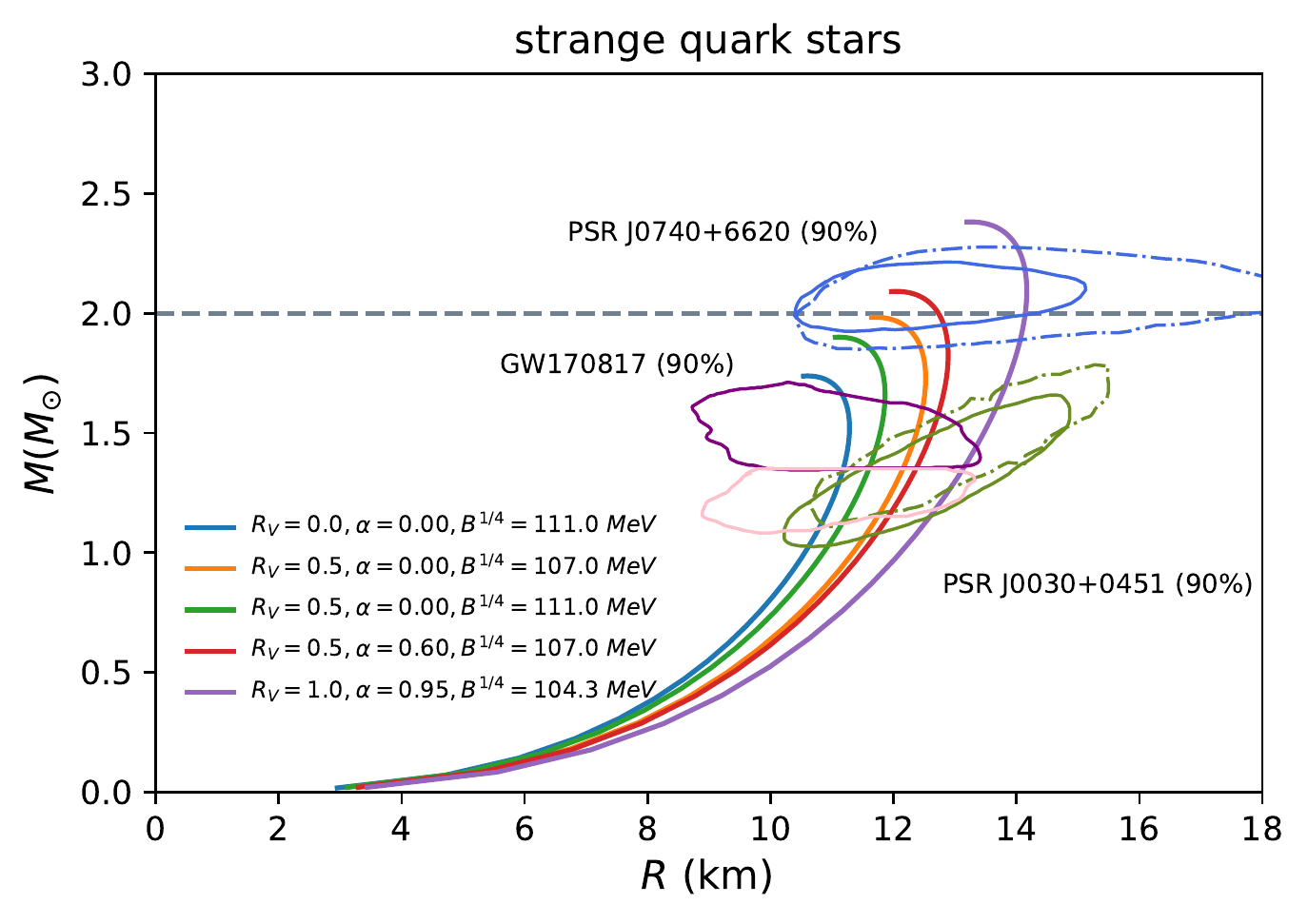}}
\caption{Mass-radius relations for nonstrange (upper panel) and strange (lower panel) quark stars in the framework of NJL-type models, by using the same sets of parameters as in Fig~\ref{fig:prho}. 
Shown together are the available mass-radius constraints from the NICER mission (PSR J0030+0451~\cite{2019ApJ...887L..24M,2019ApJ...887L..21R} and PSR J0740+6620~\cite{2021ApJ...918L..28M,2021ApJ...918L..27R}) and the binary tidal deformability constraint from LIGO/Virgo (GW170817~\cite{2017PhRvL.119p1101A,2018PhRvL.121p1101A}), at the $90\%$ confidence level. The horizontal lines in the two panels indicate $M=2\Msun$.
}    \vspace{-0.4cm}
\label{fig:star}
\end{figure}

To recap, once a particular set is selected, from Eq.~(\ref{eq17}) and Eqs.~(\ref{eq25}-\ref{eq27}), we are ready to obtain the pressure of quark matter, given as a function of the baryon chemical potential (or the baryon number density); 
Then the EoS completely determines the structure of the General Relativistic stellar models through an integration of the Tolman-Oppenheimer-Volkoff (TOV) equations~\cite{1939PhRv...55..364T,1939PhRv...55..374O}.

\subsection{Strange and nonstrange quark stars}

\begin{figure}
\centering
\vskip-2mm
\includegraphics[width=0.49\textwidth]{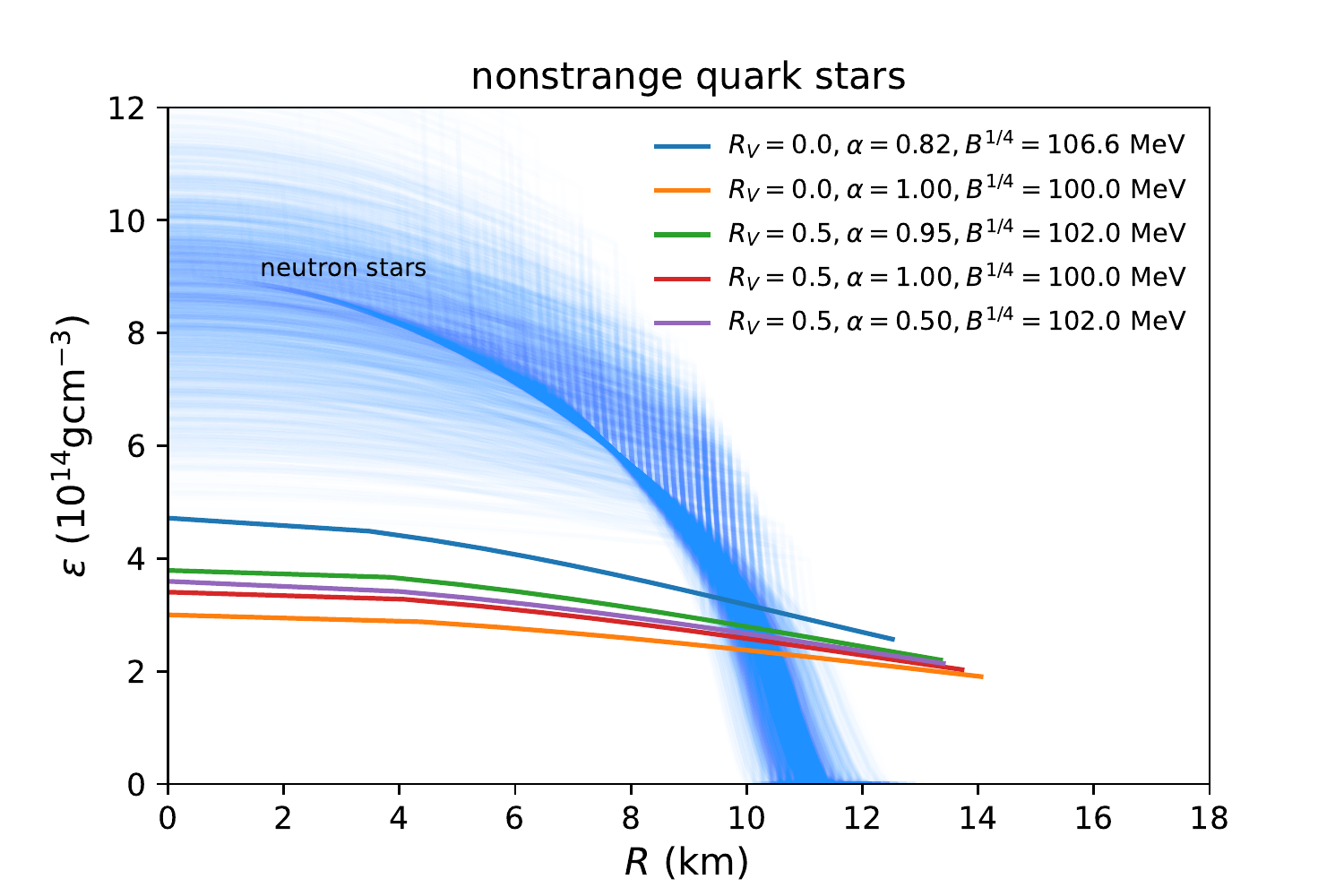}
\includegraphics[width=0.49\textwidth]{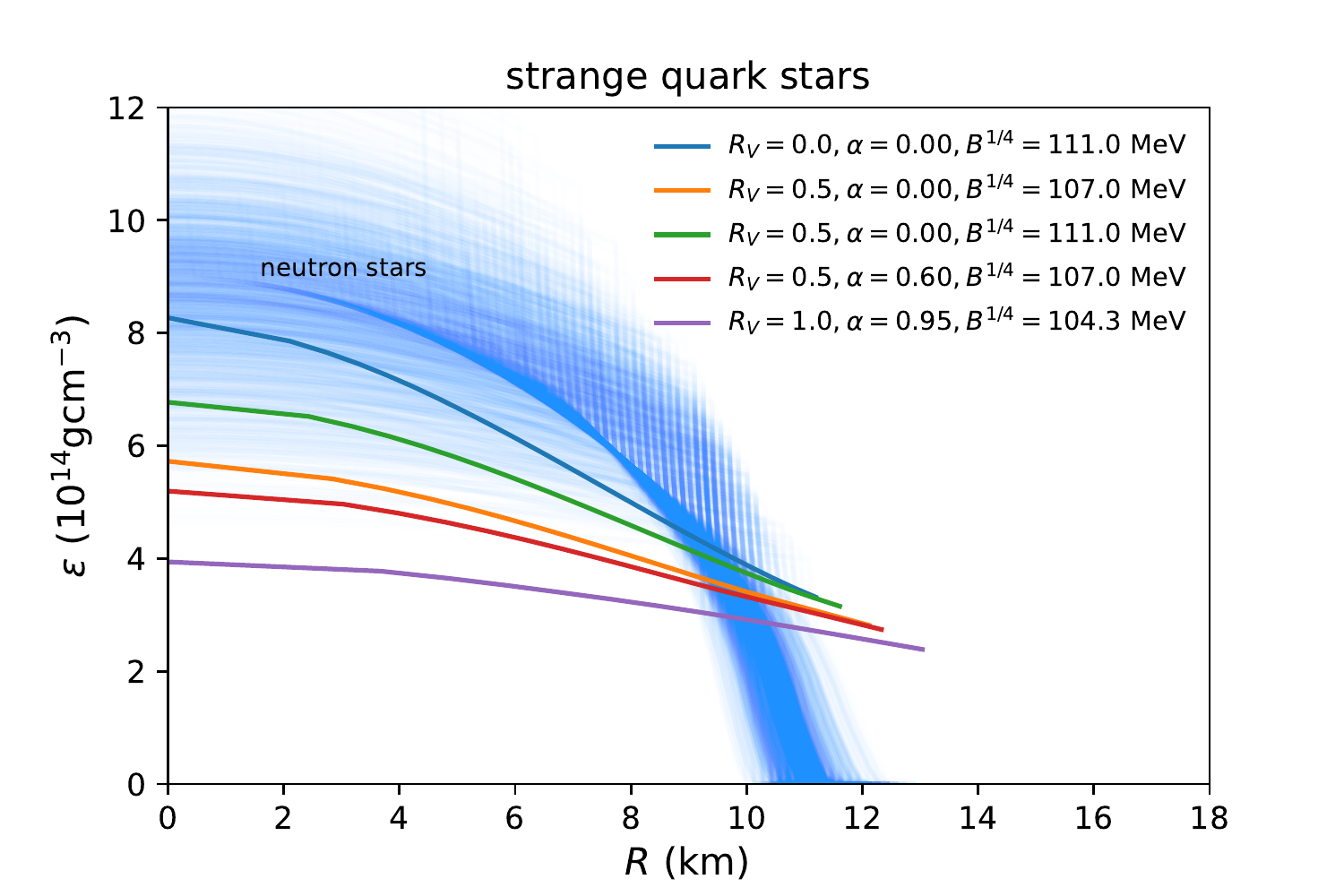}
\caption{Comparison of strange and nonstrange $1.4\Msun$ stars (calculated under the same sets of NJL parameters as in Fig~\ref{fig:prho}), as well as neutron stars~\cite{2021ApJ...913...27L} constrained by the available LIGO/Virgo and NICER data (see text for details).
}
\vspace{-0.4cm}
\label{fig:mass14}
\end{figure}

Fig.~\ref{fig:prho} shows the five EoSs for both nonstrange $ud$ matter and strange $uds$ matter obtained within the present NJL models. 
It is seen that the obtained EoSs are generally stiffer than the usual linear ones
due to that the interaction between quarks is strong within the description of the NJL-type models (see more discussions below on the sound velocity of the quark matter medium).
As mentioned in the previous section, the model parameters $(R_V, \alpha, B^{1/4})$ have been varied throughout the stability window (see Fig.~\ref{fig:stable}) to get a comprehensive picture of the corresponding self-bound stars.
For each EoS case, the central densities of the most massive stars, $\sim 3$-$5\rho_0$ for nonstrange quark stars and $\sim 4$-$7\rho_0$ for strange quark stars, are indicated with black stars, respectively.
And the EoSs of nonstrange quark stars are generally stiffer than those strange quark stars, which will be immediately reflected in the stellar mass-radius relations, as depicted in Fig.~\ref{fig:star}. 

In Fig.~\ref{fig:star}, the calculated mass-radius relations of the NJL quark stars are shown together with the constraints from both the LIGO/Virgo~\cite{2017PhRvL.119p1101A,2018PhRvL.121p1101A} and the NICER mission~\cite{2019ApJ...887L..24M,2019ApJ...887L..21R,2021ApJ...918L..28M,2021ApJ...918L..27R}.
Indeed the nonstrange quark stars are generally more massive than those strange ones: 
$M_{\rm TOV}^{\rm ud}=2.76\Msun$ vs. $M_{\rm TOV}^{\rm uds}=2.1 \Msun$ for the maximum mass.
We mention here that a maximum mass around $2.1\Msun$ was previously found for normal strange quark stars from the model calculations of the bag-model EoSs~\cite{2018PhRvD..97h3015Z} as well as the statistics analysis~\cite{2021MNRAS.506.5916L,2021ApJ...917L..22M}. 
As previously discussed, decreasing $B$, as well as increasing $\alpha$ and $R_V$, generally leads to a stiffer EoS, resulting in a larger maximum mass for quark stars. 
We see that a large set of parameters $B$, $\alpha$, and $R_V$ could in principle explain the data from LIGO/Virgo and NICER, whereas in the nonstrange case, those of very massive quark stars located on the right side of the GW170817 $90\%$ region. This discrepancy might indicate the merging sources in such a binary system are not nonstrange quark stars.
In this sense, more advanced detection techniques in the future, rendering more precise measurements for masses and radii of compact stars, hold promise for constraining their composition (see recent discussions in e.g., Refs.~\cite{2021arXiv210707979M,2021ApJ...917L..22M,2022arXiv220101217P}. 

We subsequently show in Fig.~\ref{fig:mass14} the mass density within a $1.4\Msun$ star as a function of radial coordinate, composed of entirely either nonstrange or strange quark matter. 
In the same figure, we also include the results of neutron stars constrained from combined data of LIGO/Virgo and NICER within the Bayesian statistical approach~\cite{2021ApJ...913...27L}.
One can see that the internal structure of quark stars is very different from that of neutron stars.
The quark stars, no matter strange and nonstrange ones, have a huge surface density $\sim\rho_0$, and the central density is only a few times higher than the surface one: $\lesssim2.0$ in the nonstrange case and $\lesssim2.5$ in the strange case. It contrasts the $14$ orders of magnitude difference of neutron stars.
Moreover, the more massive nonstrange quark stars, compared to the strange ones, are again reflected in their higher incompressibility.

\begin{figure*}
\resizebox*{0.48\textwidth}{0.25\textheight}
{\includegraphics[width=0.48\textwidth]{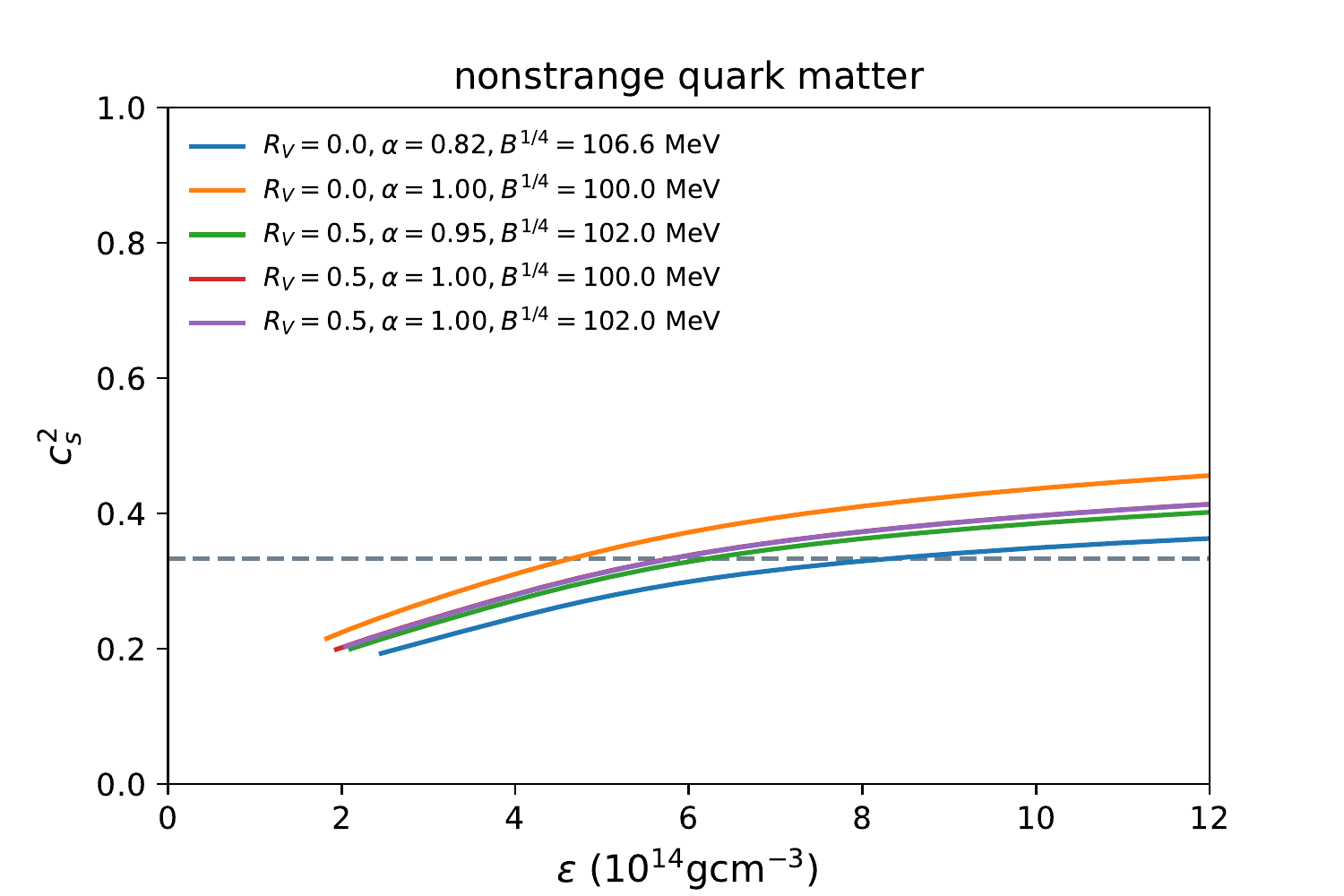}}
\resizebox*{0.48\textwidth}{0.25\textheight}
{\includegraphics[width=0.48\textwidth]{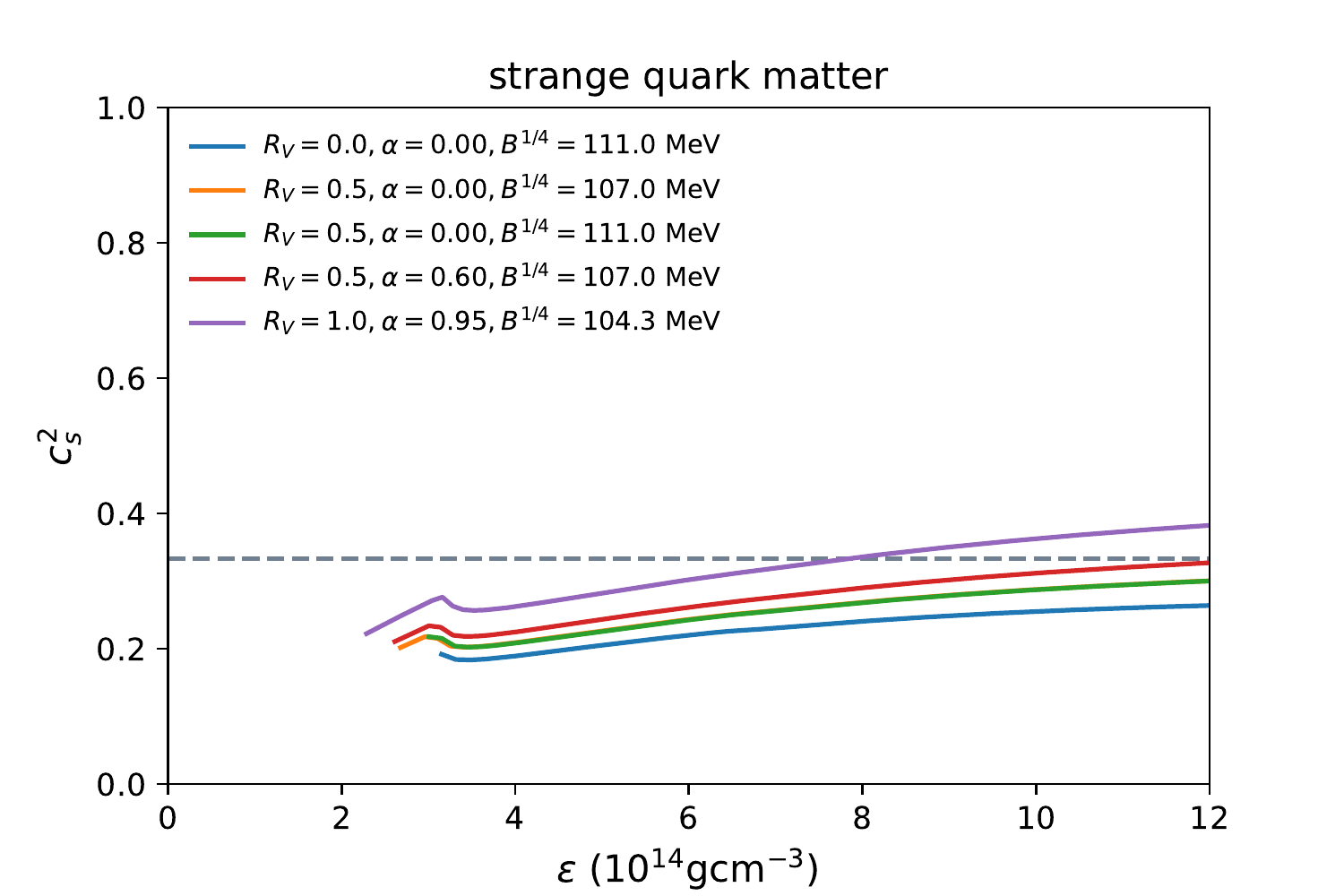}}
\caption{Sound speed squared $c_s^2$ (in units of the speed of light squared $c^2$) versus density for $ud$ and $uds$ stellar matter. The calculations are done with the same sets of parameters as in Fig~\ref{fig:prho}. 
The horizontal lines in both panels indicate the conformal limit.
}\vspace{-0.4cm}
\label{fig:cs}
\end{figure*}

We additionally calculate another important quantity of the stellar matter: 
the sound speed $c_s=\sqrt{{\mathrm{d} P}/{\mathrm{~d} \varepsilon}}$, and the sound speed squared are reported in Fig.~\ref{fig:cs} for both nonstrange and strange cases.
The conformal limit of $c_s=c/\sqrt{3}$ is indicated with the horizontal lines.
It is seen that the sound velocity shows a monotonically-increasing behavior in both nonstrange and strange cases, except there are small local maximums at the beginning of the plots in the strange cases, caused by the appearance of the strange quarks.
The conformal limit is reached earlier in the nonstrange case than in the strange case, both at about $1.5$-$3\rho_0$, then slightly exceeds the conformal limit at high densities.
Note that in the simple bag model, while keeping consistency with the observational data, the sound velocity is approximately a constant close to the conformal limit~\cite{2021ApJ...917L..22M,2021MNRAS.506.5916L}, indicating the model only accounts for weakly-interacting quarks; The current model calculations within the NJL-type models are improved in the sense that they incorporate some realistic interactions between quarks, as well as the exchange interaction channels with the mean-field approximation, beyond the basic non-perturbative phenomenon (i.e., the confinement).
Finally, in the neutron star case, the maximum sound velocity can reach $\sim0.8c$~\cite{2021ApJ...913...27L} if one takes the observational constraints into consideration (mainly the $\sim2\Msun$ TOV-mass constraint~\cite{2010Natur.467.1081D,2013Sci...340..448A,2016ApJ...832..167F,2020NatAs...4...72C,2021ApJ...915L..12F}).
Lately there are various studies~\cite{2021arXiv211014896D,2021arXiv211002100K,2021PhRvC.104e5803M,2021PhRvD.104c4011H,2021arXiv210701879L,2021ChPhC..45e5104X,2021arXiv210500029A,2021arXiv210413822M,2021PhRvD.103g1504P,2022PhRvD.105b3018T} regarding its complicated density-dependence at the density region of several times of the saturation density $\rho_0$.
We mention here that the high peak value of the sound velocity in neutron star matter~\cite{2015PhRvL.114c1103B}, compared to that in pure quark matter, can naturally be understood as a correspondence of low surface density that has been demonstrated previously in Fig.~\ref{fig:mass14}.

\section{Conclusions and summary}\label{sec:summary}

The supranuclear matter composition has long been a challenging topic that is directly related to neutron star physics.
Although we have fair good knowledge on the subnuclear matter below the neutron drip density $\varepsilon_{\rm d}\sim 10^{11}\rm g/cm^{3}$), the phase state in the stellar interior are complicated by the possible appearance of a various exotic degree of freedom, such as hyperons, kaons, Delta isobars or deconfined quarks. 
The compact stars may even be self-bound with deconfined quarks constituting the entire star, following the quantitative studies with the bag model from the seventies.

In the present study, in the framework of NJL models, based on a novel self-consistent mean field approximation by means of a Fierz transformation, we discuss the absolute stability of strange and nonstrange quark matter and compute the mass-radius relations of self-bound stars for varying vector interactions and the exchange channels.
The obtained EoSs of interacting quark matter show a non-linear (sometimes polytropic) behavior, different from those in the bag model.
Both nonstrange and strange quark stars can, in general, reconcile with the available mass and radius constraints from observational data. 
The allowed maximum mass of nonstrange stars is larger than the strange ones, up to $\sim2.7\Msun$ for an extremely low surface density close to the nuclear saturation density $\rho_0$.
The hypothetic absolute stability of quark matter, allowed by an ample parameter space in the present model calculations, not only theoretically supports quark stars as viable alternative physical model for neutron stars~\cite{1985PhLB..160..181B,1990MPLA....5.2197G,2016PhRvD..94h3010L,2018PhRvD..97h3015Z,2021PhRvL.126p2702B,2020arXiv200900942C,2021arXiv210202357T,2021arXiv210400544S,2021ApJ...922..266B}, but also could have important consequences on various astrophysics and cosmological problems, such as supernovae~\cite{1989PhRvL..63..716B,1995ApJ...440..815D,1997ApJ...481..954A}, gamma-ray bursts~\cite{1986PhRvL..57.2088A,1991ApJ...375..209H,1996ApJ...462L..63M,1996PhRvL..77.1210C,2000A&A...357..543W,2002A&A...387..725O}, fast radio bursts~\cite{2016RAA....16...80S,2018ApJ...858...88Z,2020RAA....20...56J}, pulsar glitch~\cite{1991MNRAS.250..679B,2021MNRAS.500.5336W}, cosmic rays~\cite{2003BuSSA..93.2363A}.
In future work, we plan to include the diquark channels for quark superfluidity for improving the phenomenological models of strong interactions at finite density, to advance the understanding of quark matter and make an attempt to tackle the unresolved questions in connection with it.

\medskip
\acknowledgments
We are thankful to Yan Yan, Jingyi Chao, Chengming Li, Liqun Su, Yonghui Xia, Bolin Li, and Yongfeng Huang for helpful discussions. 
The work is supported by National SKA Program of China (No.~2020SKA0120300), the National Natural Science Foundation of China (Grant No.~11873040), the science research grants from the China Manned Space Project (No. CMS-CSST-2021-B11), and the Youth Innovation Fund of Xiamen (No. 3502Z20206061).



\begin{thebibliography}{99}

\bibitem{1971PhRvD...4.1601B} A.~R. Bodmer, Phys. Rev. D \textbf{4}, 1601 (1971)
\bibitem{1984PhRvD..30..272W} E. Witten, Phys. Rev. D \textbf{30}, 272 (1984)
\bibitem{1979PhRvL..43.1292C} S.~A. Chin and A.~K. Kerman, Phys. Rev. Lett. \textbf{43}, 1292 (1979)
\bibitem[Terazawa et al.(1979)]{Terazawa1979} H. Terazawa, INS, University of Tokyo Report No. INSReport-336, 1979
\bibitem{1984PhRvD..30.2379F} E. Farhi and R.~L. Jaffe, Phys. Rev. D \textbf{30}, 2379 (1984)
\bibitem{1974PhRvD...9.3471C} A. Chodos, R.~L. Jaffe, K. Johnson, C.~B. Thorn, and V.~F. Weisskopf, Phys. Rev. D \textbf{9}, 3471 (1974)
\bibitem{2000A&A...359..311Z} J.~L. Zdunik,  Astron. Astrophys.  \textbf{359}, 311 (2000)
\bibitem{2000NuPhS..45S...1R} C.~D. Roberts and S.~M. Schmidt, Nucl. Phys. B, Proc. Suppl. \textbf{45}, S1 (2000)
\bibitem{2016EPJA...52..291C} H. Chen, J.-B. Wei, and H.-J. Schulze, Eur. Phys. J. A \textbf{52}, 291 (2016)
\bibitem{2021EPJC...81..612B} Z. Bai \& Y.-. xin . Liu, Eur. Phys. J. C \textbf{81}, 612 (2021)
\bibitem{1992RvMP...64..649K} S.~P. Klevansky, Rev. Mod. Phys. \textbf{64}, 649 (1992)
\bibitem{2005PhR...407..205B} M. Buballa, Phys. Rep. \textbf{407}, 205 (2005)
\bibitem{1994PhR...247..221H} T. Hatsuda and T. Kunihiro, Phys. Rep. \textbf{247}, 221 (1994)
\bibitem{1991PrPNP..27..195V} U. Vogl and W. Weise,  Prog. Part. Nucl. Phys.  \textbf{27}, 195 (1991)
\bibitem{1998PhLB..438..123D} M. Dey, I. Bombaci, J. Dey, S. Ray, and B.~C. Samanta, Phys. Lett. B \textbf{438}, 123 (1998)
\bibitem{1989PhLB..229..112C} S. Chakrabarty, S. Raha, and B. Sinha, Phys. Lett. B \textbf{229}, 112 (1989)
\bibitem{1999PhRvC..61a5201P} G.~X. Peng, H.~C. Chiang, J.~J. Yang, L. Li, and B. Liu, Phys. Rev. C \textbf{61}, 015201 (1999)
\bibitem{2000PhRvC..62a5204W} P. Wang, Phys. Rev. C \textbf{62}, 015204 (2000)
\bibitem{2010MNRAS.402.2715L} A. Li, R.-X. Xu, and J.-F. Lu, Mon. Notices Royal Astron. Soc.  \textbf{402}, 2715 (2010)
\bibitem{1988PhLB..200..235G} P.~A.~M. Guichon, Phys. Lett. B \textbf{200}, 235 (1988)
\bibitem{1996NuPhA.601..349G} P.~A.~M. Guichon, K. Saito, E. Rodionov, and A.~W. Thomas, Nucl. Phys. A \textbf{601}, 349 (1996)
\bibitem{1996PhLB..374...13J} X. Jin and B.~K. Jennings, Phys. Lett. B \textbf{374}, 13 (1996)
\bibitem{1997NuPhA.626..966M} H. M{\"u}ller and B.~K. Jennings, Nucl. Phys. A \textbf{626}, 966 (1997)
\bibitem{2016PhRvL.116i2501S} J.~R. Stone, P.~A.~M. Guichon, P.~G. Reinhard, and A.~W. Thomas, Phys. Rev. Lett. \textbf{116}, 092501 (2016)
\bibitem{2020JHEAp..28...19L} A. Li, et al., J. High Energy Phys. \textbf{28}, 19 (2020)
\bibitem{1996csnp.book.....G} N. Glendenning, \textit{Compact Stars. Nuclear Physics, Particle Physics and General Relativity} (Springer-Verlag, New York, 1996).
\bibitem{1999LNP...516..162M} J. Madsen, Hadrons in Dense Matter and Hadrosynthesis \textbf{516}, 162 (1999)
\bibitem{2005PrPNP..54..193W} F. Weber,  Prog. Part. Nucl. Phys.  \textbf{54}, 193 (2005)
\bibitem{2007ASSL..326.....H} P. Haensel, A.~Y. Potekhin, and D.~G. Yakovlev, Astrophysics and Space Science Library \textbf{326}, (2007)
\bibitem{2018PhRvL.120v2001H} B. Holdom, J. Ren, and C. Zhang, Phys. Rev. Lett. \textbf{120}, 222001 (2018)
\bibitem{2019PhRvD.100d3018Z} T. Zhao, et al., Phys. Rev. D \textbf{100}, 043018 (2019)
\bibitem{2019PhRvD.100l3003W} Q. Wang, C. Shi, and H.-S. Zong, Phys. Rev. D \textbf{100}, 123003 (2019)
\bibitem{2020arXiv200900942C} Z. Cao, L.-W. Chen, P.-C. Chu, and Y. Zhou, arXiv:2009.00942
\bibitem{2020MPLA...3550321W} Q. Wang, T. Zhao, and H. Zong, Mod. Phys. Lett. A \textbf{35}, 2050321 (2020)
\bibitem{2021NuPhB.97115540X} S.-S. Xu, Nucl. Phys. B. \textbf{971}, 115540 (2021)
\bibitem{2019ChPhC..43h4102W} F. Wang, Y. Cao, and H. Zong, Chin. Phys. C \textbf{43}, 084102 (2019)
\bibitem{2020PhRvD.102e4028S} L.-Q. Su, C. Shi, Y.-H. Xia, and H. Zong, Phys. Rev. D \textbf{102}, 054028 (2020)
\bibitem{2019PhRvD.100i4012Y} L.-K. Yang, X. Luo, and H.-S. Zong, Phys. Rev. D \textbf{100}, 094012 (2019)
\bibitem{2021ChPhC..45f4102W} Z.-Q. Wu, J.-L. Ping, and H.-S. Zong, Chin. Phys. C \textbf{45}, 064102 (2021)
\bibitem{2020PhRvD.101f3023L} C.-M. Li, et al., Phys. Rev. D \textbf{101}, 063023 (2020)
\bibitem{2012PhRvD..85k4017S} G.~Y. Shao, M. Colonna, M. Di Toro, B. Liu, and F. Matera, Phys. Rev. D \textbf{85}, 114017 (2012)
\bibitem{2013PhRvD..88h5001K} T. Kl{\"a}hn, R. {\L}astowiecki, and D. Blaschke, Phys. Rev. D \textbf{88}, 085001 (2013)
\bibitem{2015ApJ...810..134K} T. Kl{\"a}hn and T. Fischer, Astrophys. J. \textbf{810}, 134 (2015)
\bibitem{2017PhRvD..96h3019C} P.-C. Chu and L.-W. Chen, Phys. Rev. D \textbf{96}, 083019 (2017)
\bibitem{2018Univ....4...30C} M. Cierniak, T. Kl{\"a}hn, T. Fischer, and N.-U. Bastian, Universe \textbf{4}, 30 (2018)
\bibitem{2019JPhG...46c4002D} V. Dexheimer, R. de O. Gomes, S. Schramm, and H. Pais, J. Phys. G: Nucl. Part. Phys. \textbf{46}, 034002 (2019)
\bibitem{2020EPJST.229.3629O} K. Otto, M. Oertel, and B.-J. Schaefer, Eur. Phys. J. Spec. Top. \textbf{229}, 3629 (2020)
\bibitem{2021PhRvD.104h3011A} M.~B. Albino, R. Fariello, and F.~S. Navarra, Phys. Rev. D \textbf{104}, 083011 (2021)
\bibitem{2021Symm...13..124A} G. Alaverdyan, Symmetry \textbf{13}, 124 (2021)
\bibitem{2005PhRvC..71a5205Z} H.-S. Zong, L. Chang, F.-Y. Hou, W.-M. Sun, and Y.-X. Liu, Phys. Rev. C \textbf{71}, 015205 (2005)
\bibitem{2020PTEP.2020h3C01P} Particle Data Group, et al., Prog. Theor. Exp. Phys. \textbf{2020}, 083C01 (2020)
\bibitem{2011PhRvD..84e6010K} K. Kashiwa, T. Hell, and W. Weise, Phys. Rev. D \textbf{84}, 056010 (2011)
\bibitem{2008PhRvD..77k4028F} K. Fukushima, Phys. Rev. D \textbf{77}, 114028 (2008)
\bibitem{2008PhRvD..78c9902F} K. Fukushima, Phys. Rev. D \textbf{78}, 039902 (2008)
\bibitem{2009PhRvD..80a4015Z} Z. Zhang and T. Kunihiro, Phys. Rev. D \textbf{80}, 014015 (2009)
\bibitem{2008PhRvD..78e4001Z} H.-S. Zong and W.-M. Sun, Phys. Rev. D \textbf{78}, 054001 (2008)
\bibitem{2008IJMPA..23.3591Z} H.-S. Zong and W.-M. Sun, Int. J. Mod. Phys. A \textbf{23}, 3591 (2008)
\bibitem{2008PhRvD..77f3004P} G. Pagliara and J. Schaffner-Bielich, Phys. Rev. D \textbf{77}, 063004 (2008)
\bibitem{2012ApJ...759...57L} C.~H. Lenzi \& G. Lugones, Astrophys. J. \textbf{759}, 57 (2012)
\bibitem{2021PhRvD.103f3018Z} C. Zhang and R.~B. Mann, Phys. Rev. D \textbf{103}, 063018 (2021)
\bibitem{2021EPJC...81..921L} B.-L. Li, Y. Yan, and J.-L. Ping, Eur. Phys. J. C \textbf{81}, 921 (2021)
\bibitem{2021arXiv211209595P} A. Pfaff, H. Hansen, and F. Gulminelli, arXiv:2112.09595
\bibitem{1939PhRv...55..364T} R.~C. Tolman, Phys. Rev. \textbf{55}, 364 (1939)
\bibitem{1939PhRv...55..374O} J.~R. Oppenheimer and G.~M. Volkoff, Phys. Rev. \textbf{55}, 374 (1939)
\bibitem{2017PhRvL.119p1101A} B.~P. Abbott, et al., Phys. Rev. Lett. \textbf{119}, 161101 (2017)
\bibitem{2018PhRvL.121p1101A} B.~P. Abbott, et al., Phys. Rev. Lett. \textbf{121}, 161101 (2018)
\bibitem{2019ApJ...887L..24M} M.~C. Miller, et al., Astrophys. J. \textbf{887}, L24 (2019)
\bibitem{2019ApJ...887L..21R} T.~E. Riley, et al., Astrophys. J. \textbf{887}, L21 (2019)
\bibitem{2021ApJ...918L..28M} M.~C. Miller, et al., Astrophys. J. \textbf{918}, L28 (2021)
\bibitem{2021ApJ...918L..27R} T.~E. Riley, et al., Astrophys. J. \textbf{918}, L27 (2021)
\bibitem{2018PhRvD..97h3015Z} E.-P. Zhou, X. Zhou, and A. Li, Phys. Rev. D \textbf{97}, 083015 (2018)
\bibitem{2021MNRAS.506.5916L} A. Li, Z.-Q. Miao, J.-L. Jiang, S.-P. Tang, and R.-X. Xu, Mon. Notices Royal Astron. Soc.  \textbf{506}, 5916 (2021)
\bibitem{2021ApJ...917L..22M} Z. Miao, J.-L. Jiang, A. Li, and L.-W. Chen, Astrophys. J. \textbf{917}, L22 (2021)
\bibitem{2021arXiv210707979M} Z. Miao and A. Li, arXiv:2107.07979
\bibitem{2022arXiv220101217P} J.~P. Pereira, M. Bejger, J. Leszek Zdunik, and P. Haensel, arXiv:2201.01217
\bibitem{2021ApJ...913...27L} A. Li, Z. Miao, S. Han, and B. Zhang, Astrophys. J. \textbf{913}, 27 (2021)
\bibitem{2010Natur.467.1081D} P.~B. Demorest, T. Pennucci, S.~M. Ransom, M.~S.~E. Roberts, and J.~W.~T. Hessels, Nature \textbf{467}, 1081 (2010)
\bibitem{2013Sci...340..448A} J. Antoniadis, et al., Science \textbf{340}, 448 (2013)
\bibitem{2016ApJ...832..167F} E. Fonseca, et al., Astrophys. J. \textbf{832}, 167 (2016)
\bibitem{2020NatAs...4...72C} H.~T. Cromartie, et al., Nature Astronomy \textbf{4}, 72 (2020)
\bibitem{2021ApJ...915L..12F} E. Fonseca, et al., Astrophys. J. \textbf{915}, L12 (2021)
\bibitem{2021arXiv211014896D} C. Drischler, S. Han, and S. Reddy, arXiv:2110.14896
\bibitem{2021arXiv211002100K} T. Kojo and D. Suenaga, arXiv:2110.02100
\bibitem{2021PhRvC.104e5803M} J. Margueron, H. Hansen, P. Proust, and G. Chanfray, Phys. Rev. C \textbf{104}, 055803 (2021)
\bibitem{2021PhRvD.104c4011H} M. Hippert, E.~S. Fraga, and J. Noronha, Phys. Rev. D \textbf{104}, 034011 (2021)
\bibitem{2021arXiv210701879L} H.~K. Lee, Y.-L. Ma, W.-G. Paeng, and M. Rho, arXiv:2107.01879
\bibitem{2021ChPhC..45e5104X} C. Xia, Z. Zhu, X. Zhou, and A. Li, Chin. Phys. C \textbf{45}, 055104 (2021)
\bibitem{2021arXiv210500029A} S. Anti{\'c}, M. Shahrbaf, D. Blaschke, and A.~G. Grunfeld, arXiv:2105.00029
\bibitem{2021arXiv210413822M} Y.-L. Ma and M. Rho, arXiv:2104.13822
\bibitem{2021PhRvD.103g1504P} R.~D. Pisarski, Phys. Rev. D \textbf{103}, L071504 (2021)
\bibitem{2022PhRvD.105b3018T} H. Tan, T. Dore, V. Dexheimer, J. Noronha-Hostler, and N. Yunes, Phys. Rev. D \textbf{105}, 023018 (2022)
\bibitem{2015PhRvL.114c1103B} P. Bedaque \& A.~W. Steiner, Phys. Rev. Lett. \textbf{114}, 031103 (2015)
\bibitem{1985PhLB..160..181B} G. Baym, E.~W. Kolb, L. McLerran, T.~P. Walker, and R.~L. Jaffe, Phys. Lett. B \textbf{160}, 181 (1985)
\bibitem{1990MPLA....5.2197G} N.~K. Glendenning, Mod. Phys. Lett. A \textbf{5}, 2197 (1990)
\bibitem{2016PhRvD..94h3010L} A. Li, et al., Phys. Rev. D \textbf{94}, 083010 (2016)
\bibitem{2021PhRvL.126p2702B} I. Bombaci, A. Drago, D. Logoteta, G. Pagliara, and I. Vida{\~n}a, Phys. Rev. Lett. \textbf{126}, 162702 (2021)
\bibitem{2021arXiv210202357T} S. Traversi, P. Char, G. Pagliara, and A. Drago, arXiv:2102.02357
\bibitem{2021arXiv210400544S} J. Sedaghat, S.~M. Zebarjad, G.~H. Bordbar, B. Eslam Panah, and R. Moradi, arXiv:2104.00544
\bibitem{2021ApJ...922..266B} Z. Bai, W.-. jie . Fu, and Y.-. xin . Liu, Astrophys. J. \textbf{922}, 266 (2021)
\bibitem{1989PhRvL..63..716B} O.~G. Benvenuto and J.~E. Horvath, Phys. Rev. Lett. \textbf{63}, 716 (1989)
\bibitem{1995ApJ...440..815D} Z. Dai, Q. Peng, and T. Lu, Astrophys. J. \textbf{440}, 815 (1995)
\bibitem{1997ApJ...481..954A} J.~D. Anand, A. Goyal, V.~K. Gupta, and S. Singh, Astrophys. J. \textbf{481}, 954 (1997)
\bibitem{1986PhRvL..57.2088A} C. Alcock, E. Farhi, and A. Olinto, Phys. Rev. Lett. \textbf{57}, 2088 (1986)
\bibitem{1991ApJ...375..209H} P. Haensel, B. Paczynski, and P. Amsterdamski, Astrophys. J. \textbf{375}, 209 (1991)
\bibitem{1996ApJ...462L..63M} F. Ma and B. Xie, Astrophys. J. \textbf{462}, L63 (1996)
\bibitem{1996PhRvL..77.1210C} K.~S. Cheng and Z.~G. Dai, Phys. Rev. Lett. \textbf{77}, 1210 (1996)
\bibitem{2000A&A...357..543W} X.~Y. Wang, Z.~G. Dai, T. Lu, D.~M. Wei, and Y.~F. Huang,  Astron. Astrophys.  \textbf{357}, 543 (2000)
\bibitem{2002A&A...387..725O} R. Ouyed and F. Sannino,  Astron. Astrophys.  \textbf{387}, 725 (2002)
\bibitem{2016RAA....16...80S} Z. Shand, A. Ouyed, N. Koning, and R. Ouyed, Research in  Astron. Astrophys.  \textbf{16}, 80 (2016)
\bibitem{2018ApJ...858...88Z} Y. Zhang, J.-J. Geng, and Y.-F. Huang, Astrophys. J. \textbf{858}, 88 (2018)
\bibitem{2020RAA....20...56J} J.-C. Jiang, et al., Research in  Astron. Astrophys.  \textbf{20}, 056 (2020)
\bibitem{1991MNRAS.250..679B} O.~G. Benvenuto and J.~E. Horvath, Mon. Notices Royal Astron. Soc.  \textbf{250}, 679 (1991)
\bibitem{2021MNRAS.500.5336W} W.~H. Wang, et al., Mon. Notices Royal Astron. Soc.  \textbf{500}, 5336 (2021)
\bibitem{2003BuSSA..93.2363A} D.~P. Anderson, The Bulletin of the Seismological Society of America \textbf{93}, 2363 (2003)


\bibliographystyle{apsrev4-1}
\end{thebibliography}
\end{document}